\documentclass{article}

\usepackage{arxiv}

\usepackage[utf8]{inputenc} %
\usepackage[T1]{fontenc}    %
\usepackage{hyperref}       %
\usepackage{url}            %
\usepackage{booktabs}       %
\usepackage{amsfonts}       %
\usepackage{nicefrac}       %
\usepackage{microtype}      %
\usepackage{lipsum}		%
\usepackage{graphicx}
\usepackage{natbib}
\usepackage{doi}
\usepackage{array}

\usepackage{amsmath, amssymb, amsfonts, amsthm}
\usepackage{bm}
\usepackage{placeins} %

\DeclareMathOperator*{\argmin}{arg\,min}

\newcommand\ci{\perp\!\!\!\perp}

\title{Adaptive Nonlinear Data Assimilation \linebreak through P-Spline Triangular Measure Transport}

\author{ 
	\href{https://orcid.org/0000-0002-6220-5680}{\includegraphics[scale=0.06]{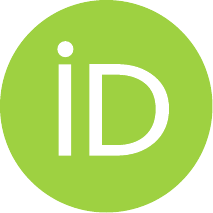}\hspace{1mm}Berent Å.~S.~Lunde}$^{1,2}$\\[2mm]
	$^1$Department of Informatics\\
	University of Bergen\\
	Bergen, Norway \\
    $^2$Equinor, Bergen, Norway\\
	\texttt{berent.a.lunde@uib.no} \\
    \And
\href{https://orcid.org/0000-0003-3508-6214}{\includegraphics[scale=0.06]{orcid.pdf}\hspace{1mm}Maximilian Ramgraber}$^{3*}$ \\
	$^3$Department of Geoscience \& Engineering\\
	Delft University of Technology\\
	Delft, 2628 CN \\
	\texttt{m.ramgraber@tudelft.nl} \\
    *Corresponding author \\
}

\renewcommand{\shorttitle}{\textit{arXiv} Template}

\hypersetup{
pdftitle={A template for the arxiv style},
pdfsubject={q-bio.NC, q-bio.QM},
pdfauthor={Berent Å.~S.~Lunde, Maximilian~Ramgraber},
pdfkeywords={data assimilation, triangular measure transport, adaptation},
}

\begin{document}
\maketitle
\begin{abstract}

Non-Gaussian statistics are a challenge for data assimilation. Linear methods oversimplify the problem, yet fully nonlinear methods are often too expensive to use in practice. The best solution usually lies between these extremes. Triangular measure transport offers a flexible framework for nonlinear data assimilation. Its success, however, depends on how the map is parametrized. Too much flexibility leads to overfitting; too little misses important structure. To address this balance, we develop an adaptation algorithm that selects a parsimonious parametrization automatically. Our method uses P-spline basis functions and an information criterion as a continuous measure of model complexity. This formulation enables gradient descent and allows efficient, fine-scale adaptation in high-dimensional settings. The resulting algorithm requires no hyperparameter tuning. It adjusts the transport map to the appropriate level of complexity based on the system statistics and ensemble size. We demonstrate its performance in nonlinear, non-Gaussian problems, including a high-dimensional distributed groundwater model.

\end{abstract}


\section{Introduction}

Ensemble data assimilation is an important tool for statistical inference in high-dimensional dynamical systems. It combines model predictions with observations to estimate uncertain states and parameters, and it is widely used in geoscience, engineering, and environmental modeling \citep[e.g., ][]{carrassi2018data,fletcher2017data}.

In many applications, forward models are nonlinear and prior distributions are non-Gaussian. Consequently, posterior distributions may exhibit skewness, multimodality, and nonlinear dependence structure.
Classical ensemble methods, including ensemble filters and smoothers \citep{evensen2003ensemble,evensen2000ensemble} and iterative ensemble smoothers \citep{emerick2013ensemble,white2018model}, usually rely on linear updates and repeated forward model evaluations.
While effective in near linear-Gaussian settings, these methods do not in general produce reliable posterior samples and may require multiple sequential simulations of the forward model.

Triangular transport maps provide a promising framework for nonlinear sampling-based inference \citep{spantini2022coupling}. A triangular transport map constructs a bijection between a target distribution of interest and a simple user-defined reference measure. Once learned, it enables target density evaluations, unconditional sampling, and conditional sampling without additional forward model runs. In nonlinear settings, transport maps can characterize complex posterior structure while retaining computational tractability.

The success of triangular transport depends critically on controlling map complexity. Overly restrictive maps reproduce the limitations of linear–Gaussian approximations and fail to capture non-Gaussian structure. By contrast, overly nonlinear updates may overfit to sampling noise and incur unfavourable bias-variance tradeoffs. This challenge has motivated the development of adaptation strategies balancing approximation accuracy and statistical stability.

Existing adaptation approaches are based on greedy selection over discrete model classes or structurally constrained Bayesian formulations. 
Greedy strategies such as \citet{baptista2024representation} incrementally expand the function space and refit the map at each stage, which can become computationally expensive in high dimensions. Bayesian structural approaches such as \citet{katzfuss2024scalable} introduce sparsity and scalability (in particular for settings like random fields) through prior assumptions on the map, but restrict the functional form of the transformation.

In this work, we develop a continuous complexity adaptation framework for triangular transport maps. We parametrise map components within a family of functions indexed by a smoothness parameter and formulate complexity selection as the optimisation of this smoothing parameter over an information criterion. The resulting adaptation objective admits efficient gradient-based exploration of the smoothing parameter, enabling practical application in high-dimensional ensemble settings.

The remainder of the paper is organised as follows. Section~\ref{sec:background} reviews sampling-based inference and ensemble data assimilation with triangular transport maps. Section~\ref{sec:methodology}
introduces the proposed gradient-based adaptation framework. Section~\ref{sec:examples} presents numerical examples, including a distributed groundwater flow model. Section~ \ref{sec:conclusions} concludes.

\section{Sampling based inference through triangular transport}
\label{sec:background}
The following section introduces transport of measure using triangular transport maps in the context of sampling based Bayesian inference and ensemble data assimilation \citep{spantini2022coupling}.
For a general introduction and overview of the capabilities of triangular transport, we refer to \citet{ramgraber_friendly_2025}.

\subsection{Change of measure}

Let $x \in \mathbb{R}^d$ be a continuous random variable with density $\pi(x)$, which we refer to as the \emph{target density}. We assume that $\pi$ is not available in closed form, but that we have access to independent samples $\{x^{(i)}\}_{i=1}^n$ drawn from $\pi$.

Let $S:\mathbb{R}^d \rightarrow \mathbb{R}^d$ be a continuously differentiable and invertible map. The map $S$ transforms $x$ into a new random variable
\begin{align}
z = S(x),    
\end{align}
which we require to follow a prescribed \emph{reference density} $\eta(z)$. Throughout this work, we take $\eta$ to be the multivariate standard Gaussian, which is a product measure with independent components and admits efficient sampling.

Because $S$ is invertible, it induces a change of measure between $\pi$ and $\eta$. Specifically, the change-of-variables formula gives
\begin{align}
\pi(x)
&=
\eta\!\left(S(x)\right)
\left| \det \nabla S(x) \right|,
\\
\eta(z)
&=
\pi\!\left(S^{-1}(z)\right)
\left| \det \nabla S^{-1}(z) \right|.
\end{align}

If $S$ exactly transports $\pi$ to $\eta$, then the pushforward of $\pi$ under $S$ equals $\eta$, which we denote by
\begin{align}
S_{\#} \pi = \eta.    
\end{align}
Conversely, the inverse map $S^{-1}$ pulls $\eta$ back to $\pi$, denoted via $S^{\#}\eta=\pi$

In practice, we construct an approximation $S$ from a finite ensemble. The induced density, $S^{\#}\eta$, then serves as an approximation to the unknown target density $\pi$. We refer this as the \emph{pullback density} induced by $S$, while the density of $S(x)$ defines the corresponding \emph{pushforward density}. When $S$ approximates the exact transport, these induced densities approximate $p$ and $\eta$, respectively.

Our objective is therefore to construct a map $S$ that transports the ensemble distribution as closely as possible to the reference measure. Once such a map is available, the change-of-measure relationship provides approximate density evaluation of the target and enables sampling from the induced pullback distribution via the inverse map.

\subsection{Triangular maps}

The change-of-measure formulation does not uniquely determine the transport map $S$. To obtain a structured/modular(?) and computationally tractable representation, we adopt a triangular construction known as the Knothe-Rosenblatt rearrangement \citep{rosenblatt1952remarks}.

The map $S$ is (lower) triangular if
\begin{align}
S(x)
=
\begin{bmatrix*}[l]
S_1(x_1) \\
S_2(x_1, x_2) \\
\vdots \\
S_d(x_1, \ldots, x_p)
\end{bmatrix*},
\end{align}
so that each component $S_j:\mathbb{R}^{j}\mapsto\mathbb{R}$ depends only on the first $j$ variables. If each $S_j$ is strictly increasing in its last argument $x_j$, the map is bijective and continuously invertible.

The Jacobian of a triangular map is lower triangular, and its determinant simplifies to
\begin{align}
\det \nabla S(x)
=
\prod_{j=1}^d
\frac{\partial S_j(x_1,\ldots,x_j)}{\partial x_j}.
\end{align}
Thus the log-determinant reduces to a sum of one-dimensional (well behaved log) derivatives, which is essential for scalable learning and inference.

Because we choose the reference $\eta$ to be the multivariate standard Gaussian, its components are independent. Sparsity in the triangular map therefore corresponds directly to conditional independence structure in the target distribution. This relationship enables high-dimensional problems to be decomposed into a sequence of low-dimensional transformations \citep{spantini2018inference}, which is particularly advantageous in ensemble data assimilation \citep{lunde2025ensemble}. This relationship also places an importance on the ordering of $x$, as this matters for conditional independence (and thus the sparsity of $S_j$).

\subsection{Sampling the conditional and ensemble data assimilation}

A key advantage of triangular transport maps is that they enable efficient conditional sampling when the reference measure is a product of marginal distributions. 

Partition the state as $x=(x_a,x_b)$, where $x_a^*$ is held fixed (e.g. conditioned upon) and $x_b$ conditioned on $x_a^*$ is to be sampled. A triangular map can be written in block form as
\begin{align}
S(x)
=
\begin{bmatrix*}[l]
S_a(x_a)\\
S_b(x_a,x_b)
\end{bmatrix*},
\end{align}
where the block $S_b$ remains triangular in $x_b$.

Because the reference measure factorises as
\begin{align}
    \eta(z)=\eta_a(z_a)\eta_b(z_b),
\end{align}
conditional sampling can be performed directly in the reference space.
Given $x_a^*$, one may draw
\begin{align}
    z_b \sim \eta_b,    
\end{align}
and then recover an updated $x_b$ by solving the triangular system
\begin{align}
x_b|x_a^* = 
S_b^{-1}(x_a^*,z_b),
\end{align}
through sequential one dimensional solves, $S_b(x_a^*,x_b)=z_b$, due to monotonicity in each triangular component. This yields samples from the conditional pullback distribution approximating $\pi(x_b|x_a)$ induced by the fitted map $S$.

In ensemble data assimilation, one typically retains and updates existing ensemble members rather than drawing new samples. Given an ensemble $\{(x_a^{(i)},x_b^{(i)})\}_{i=1}^n$ and observation $x_a^*$, we compute the block pushforward
\begin{align}
z_b^{(i)} = S_b\bigl(x_a^{(i)},x_b^{(i)}\bigr).
\end{align}
The update then replaces the conditioned variables by $x_a^*$ while retaining each member's latent coordinate $z_b^{(i)}$, and maps back through the inverse block map
\begin{align}
    x_b^{(i)} = S_b^{-1}(x_a^*,\cdot) \circ S_b\bigl(x_a^{(i)},x_b^{(i)}\bigr).
\end{align}
This again occurs in practice by one-dimensional solves for $x_b^{(i)}$
\begin{align}
S_b(x_a^*,x_b^{(i)})=z_b^{(i)},
\end{align}
for each ensemble member $i$.

Thus, triangular transport maps provide a direct and computationally efficient mechanism for ensemble updates: conditioning is carried out in independent reference/Gaussian coordinates, while the triangular inverse ensures consistent transformation back to state space. Notice also how only the lower block $S_b$ is necessary for conditioning updates.

\subsection{Learning the map from the ensemble}

Given a finite ensemble $\{x^{(i)}\}_{i=1}^n \sim \pi$, our goal is to construct a triangular map $S$ that transports the target distribution to the reference. When $S$ exactly satisfies $S_\# \pi = \eta$, the induced pullback density coincides with $\pi$. In practice, this transport must be approximated from the finite ensemble.

For triangular maps, learning can be performed component-wise. The $j$-th component of $S_j(x_1,\ldots,x_j)$ is obtained by minimizing the relative Kullback--Leibler divergence between the induced pullback density and the reference \citep{marzouk2016introduction}. This yields the population objective
\begin{align}
\mathcal{J}_j(S_j;\pi)
=
\mathbb{E}_{\pi}
\left[
\frac{1}{2} S_j(x_1,\ldots,x_j)^2
-
\log
\frac{\partial S_j(x_1,\ldots,x_j)}{\partial x_j}
\right].
\label{eq:triangular-transport-ideal-objective}
\end{align}
Since both are unknown, the expectation is replaced by a Monte Carlo approximation over the ensemble,
\begin{align}
\mathcal{J}_j(S_j;\pi^*)
=
\frac{1}{n}
\sum_{i=1}^n
\left[
\frac{1}{2} S_j(x^{(i)}_1,\ldots,x^{(i)}_j)^2
-
\log
\frac{\partial S_j(x^{(i)}_1,\ldots,x^{(i)}_j)}{\partial x^{(i)}_j}
\right].
\label{eq:triangular-transport-objective}
\end{align}
In which $\pi^*$ is the empirical distribution from the ensemble.
Minimising this empirical objective yields a maximum-likelihood estimator of the map parameters under the induced model.

Early algorithms for triangular transport, such as Algorithm~1 in Section~4.3 of \citet{baptista2024representation}, construct maps by selecting basis functions from a discrete set of candidate complexities in a greedy manner. Complexity is increased stepwise until improvements in the objective, measured by cross-validation \citep{stone1974cross}, become negligible.

Alternative approaches, such as \citet{katzfuss2024scalable}, adopt structurally constrained transport maps with Bayesian priors on their parameters, and infer map coefficients through posterior updating instead of likelihood-based KL minimization.

These methods highlight the importance of controlling map complexity when learning from finite ensembles. In the next section, we introduce a continuous adaptation strategy that replaces discrete complexity levels by a smoothing parameter optimised directly with respect to an information criterion.

\section{Gradient-based exploration of triangular map complexity}
\label{sec:methodology}

This section introduces an adaptive algorithm for selecting the complexity of triangular transport map components. In ensemble-based data assimilation, map components must be flexible enough to capture non-Gaussian structure, yet simple enough to avoid overfitting finite ensemble noise. Striking this balance is critical, particularly in high-dimensional systems where each component may depend on many others.

We develop a computationally efficient procedure that adjusts map complexity automatically. The key idea is to measure model fit and model complexity separately, using an information criterion that penalizes excessive flexibility. Rather than searching over a discrete set of candidate models \citep{baptista2024representation}, we embed map components in a continuously indexed family of functions whose effective complexity can be tuned smoothly.

This continuous parametrisation enables gradient-based optimisation of the optimal complexity level. The resulting algorithm adapts each map component using principled model selection while remaining scalable to high-dimensional assimilation problems.

\subsection{Balancing flexibility and robustness in transport maps}
\begin{figure}[ht!]
  \centering
  \includegraphics[width=\textwidth]{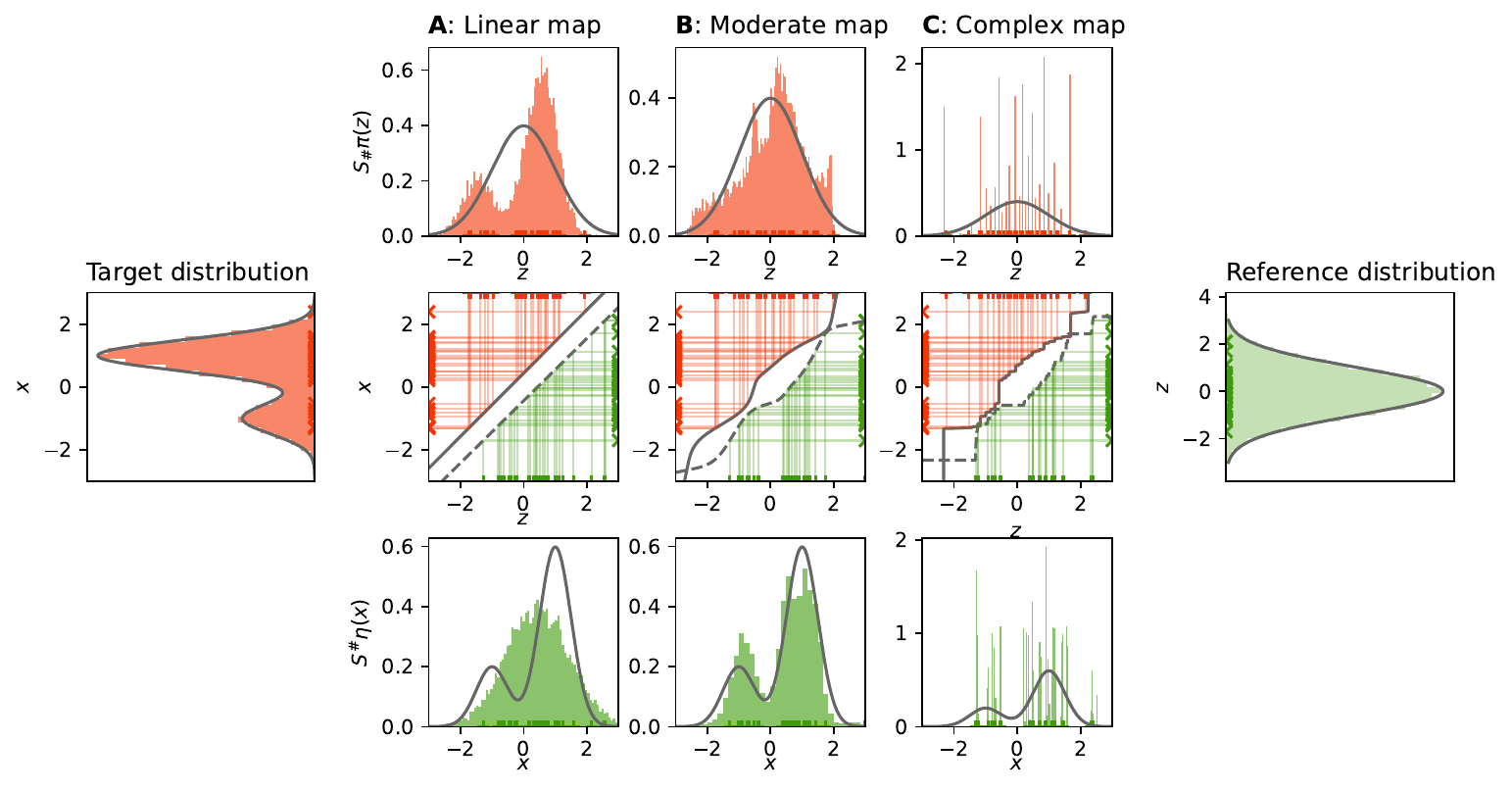}
  \caption{Effect of map complexity in a univariate transport example. A bimodal prior target density (left) is approximated from a finite ensemble. Three monotone map components of increasing flexibility (middle: A, B, C) are fitted. Using the forward map (\textcolor{custom_orange}{orange}, grey filled line in center row) pushes samples to a Gaussian reference (top row). Insufficient flexibility leaves residual non-Gaussian structure (A, top), while excessive flexibility overfits ensemble fluctuations and yields unstable pushforward $S_\#\pi$ (C, top). An intermediate level of complexity achieves stable Gaussianization (B, top). Likewise, an inverse map (\textcolor{custom_green}{green}, grey dashed line in center row) of insufficient flexibility fails to convert a standard Gaussian reference (right) to a non-Gaussian target (A, bottom), while excessive flexibility overfits ensemble fluctuations and yields unstable pullbacks $S^\#\eta$ (C, bottom).}

  \label{fig:map complexity importance}
\end{figure}
Triangular transport maps are learned from finite prior ensembles. Each map component must be flexible enough to capture non-Gaussian structure, yet simple enough to avoid fitting sampling noise. The appropriate level of complexity depends on the information content of the ensemble.

Figure \ref{fig:map complexity importance} illustrates this tradeoff in a univariate example. A map that is too simple fails to remove non-Gaussian structure. A highly flexible map interpolates ensemble fluctuations and behaves unstably between ensemble members. An intermediate level of flexibility produces a stable Gaussianizing transformation.

Overfitting in this setting can be understood as a lack of stability under resampling. If an independent ensemble were drawn from the same prior distribution, an overfitted map would not reliably transport it to the reference. A useful transport map should approximate the population transformation implied by the prior distribution, not the specific realisation of the ensemble.

The map is estimated by minimising the empirical negative log-likelihood in Equation~\eqref{eq:triangular-transport-objective}, which is a Monte Carlo approximation of the population objective in Equation~\eqref{eq:triangular-transport-ideal-objective}.
Increasing parameter flexibility always reduces this empirical objective. However, because the expectation over $\pi$ is replaced by a finite ensemble average, the minimised empirical objective underestimate the true population objective. This optimistic bias grows with model flexibility and leads to overfitting.

To control this effect, we separate model fit from model complexity using an information criterion. We adopt the Akaike Information Criterion (AIC) \citep{akaike1974new},
\begin{align}
    \texttt{AIC} = 2~\texttt{nll} + 2~\texttt{edf}
\end{align}
where \texttt{nll} is the minimised negative log-likelihood and \texttt{edf} denotes the effective degrees of freedom. The AIC approximates expected performance under independent resampling \citep{burnham2003model}, i.e. evaluation of the learnt map under the population objective in Equation~\eqref{eq:triangular-transport-ideal-objective}, and therefore directly targets the stability requirement described above. Unlike the cross-validation in \citet{baptista2024representation}, it can be evaluated from a single fitted model and is therefore suitable for high-dimensional data assimilation problems. In our implementation, we will use the corrected AIC (AICc) proposed by \citet{sugiura1978further, hurvich1989regression}, which displays greater robustness to small ensemble-size regimes in regression settings.

In the next subsection, we introduce a continuously parametrised family of map components whose effective degrees of freedom can be adjusted smoothly, enabling direct optimisation of model complexity.

\subsection{Continuously controlled map complexity via P-splines}
\begin{figure}[ht!]
  \centering
  \includegraphics[width=\textwidth]{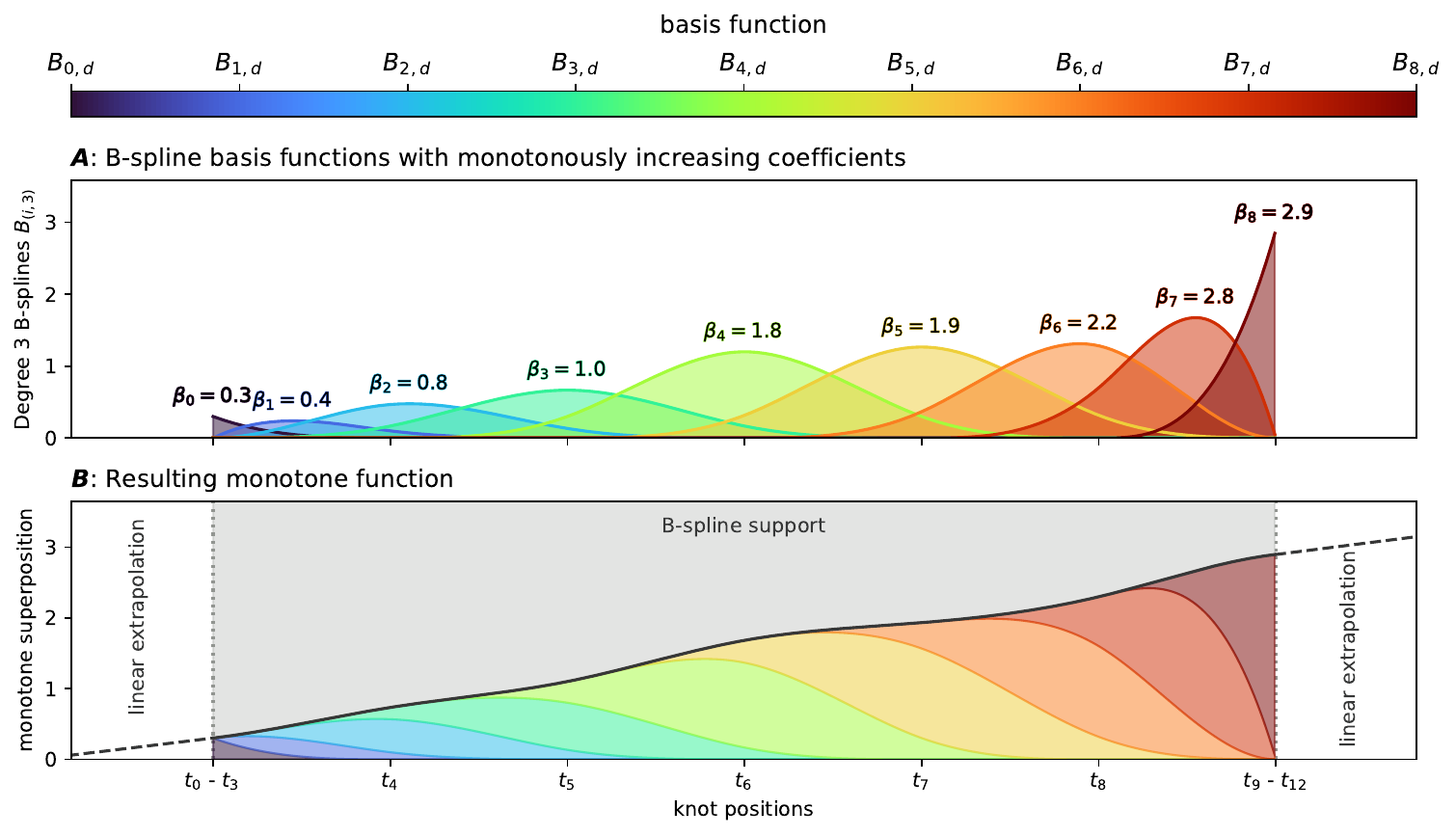}
  \caption{Assigning monotonously increasing coefficients to each B-spline basis function $B_{i,d}$ (A) results in a monotone function from the superposition $f(x,\boldsymbol{\beta})$ of the scaled basis functions (B). Linearly extrapolating the B-splines furthermore ensures monotonicity beyond the support of the B-spline.}
  \label{fig:B_splines_monotone}
\end{figure}

To enable principled complexity control, we embed each map component $S_j$ in a continuously parametrised spline family. Rather than selecting basis size discretely, flexibility is governed by a smoothing parameter that 
adjusts effective degrees of freedom smoothly.

We adopt a separable additive formulation,
\begin{align}
S_j(x_1,\ldots,x_j)
=
\sum_{k \in \mathcal{P}_j} s_{jk}(x_k),
\end{align}
where $\mathcal{P}_j$ denotes the parent set in the triangular structure. Each univariate term is represented using a B-spline basis,
\begin{align}
s_{jk}(x_k)
=
\sum_{m} \beta_m \, B_m(x_k),
\end{align}
where $\{B_m\}$ are B-spline basis functions and $\beta$ denotes the coefficient vector.

Monotonicity in the final argument $x_j$ is enforced through a reparameterisation of the spline coefficients that guarantees nonnegative partial derivatives. In practice, this is achieved by constraining successive coefficients to be monotonically increasing, as illustrated in Figure~\ref{fig:B_splines_monotone}. This construction ensures invertibility of the triangular map component while retaining a linear representation in the underlying coefficients. Further implementation details are provided in Appendix~\ref{appendix:p-splines}.

Let
\begin{align}
\mathcal{L}_j(\beta)
:=
\mathcal{J}_j\!\left(S_j(\beta); \pi^*\right)
\end{align}
denote the empirical objective in Equation~\eqref{eq:triangular-transport-objective} expressed in terms of the spline coefficients. We introduce a quadratic roughness penalty and define the penalized objective
\begin{align}
\mathcal{L}_{j,\lambda}(\beta)
:=
\mathcal{L}_j(\beta)
+
\beta^\top \Lambda D^\top D \beta,
\end{align}
where $D$ is a block diagonal difference matrix acting on the spline coefficients and $\Lambda=\mathrm{diag}(\lambda)$ collects smoothing parameters associated with the different spline components. The blocks of $D$ correspond to the monotone and non-monotone parts of the map component $S_j$. Monotonicity of the constrained components is enforced through the shape-constrained P-spline construction of \citet{pya2015shape}.

The penalty serves two purposes. First, it encodes an explicit preference for smooth functions, stabilising behaviour between ensemble members and reducing overfitting in regions with limited data support. Second, it introduces a continuous complexity parameter $\lambda$, replacing discrete model selection over basis size.

For fixed $\lambda$, coefficients are obtained from the \emph{inner optimisation problem}
\begin{align}\label{eq:inner objective}
\hat{\beta}(\lambda)
=
\arg\min_{\beta}
\mathcal{L}_{j,\lambda}(\beta).
\end{align}
For penalised likelihood models, \citet{wood2020inference} gives the effective degrees of freedom of the fitted component are given by
\begin{align}\label{eq:effective degrees of freedom}
\texttt{edf}_j(\lambda)
=
\mathrm{tr}
\left[
\left(\nabla_\beta^2 \mathcal{L}_{j,\lambda}(\hat{\beta})\right)^{-1}
\nabla_\beta^2 \mathcal{L}_j(\hat{\beta})
\right].
\end{align}
As $\lambda \to 0$, the effective degrees of freedom approach the full basis dimension $K$. As $\lambda \to \infty$, they shrink toward the dimension of the null space of the penalty matrix $D^\top D$, which is 2 for each sub-block \citep{wood2020inference}.

Thus, following the adoption of the AIC, we define the objective for the \emph{outer optimisation problem}
\begin{align}\label{eq:outer objective}
\mathcal{A}_j(\lambda)
:=
\,\mathcal{L}_j(\hat{\beta}(\lambda))
+
\,\texttt{edf}_j(\lambda),
\end{align}
(ignoring the constant factor 2) and seek
\begin{align}
\lambda^*
=
\arg\min_{\lambda \ge 0}
\mathcal{A}_j(\lambda).
\end{align}

This formulation converts discrete model selection into continuous optimisation over a scalar smoothing parameter, enabling scalable and automated complexity adaptation across many transport map components.

\subsection{Efficient gradient-based adaptation of complexity}
\label{subsec:gradient based adaptation}
The outer objective Equation~\eqref{eq:outer objective} depends on the smoothing parameter $\lambda$ only through the minimiser $\hat{\beta}$ of the inner objective Equation~\eqref{eq:inner objective}.
A naive strategy would evaluate $\mathcal{A}_j(\lambda)$ on a grid of candidate values. However, each evaluation requires solving the inner optimisation problem. In high-dimensional transport maps with many components, repeated refitting is computationally prohibitive.

Instead, we compute the gradient of the outer objective directly to facilitate efficient gradient based exploration.
Because $\hat{\beta}$ depends implicitly on $\lambda$, differentiation of the outer objective requires implicit differentiation. For clarity here, we define the two-argument outer objective
\begin{align}
    \tilde{\mathcal{A}}_j(\beta, \lambda)
    := \mathcal{L}_j(\beta) + \texttt{edf}_j(\beta, \lambda)
\end{align}
and the reduced objective in Equation~\eqref{eq:outer objective} is then
\begin{align}
    \mathcal{A}_j(\lambda) = \tilde{\mathcal{A}}_j(\hat{\beta}(\lambda), \lambda),
\end{align}
where $\hat{\beta}(\lambda)$ is defined by Equation~\eqref{eq:inner objective}.

At the optimum, the first-order condition holds:
\begin{align}
    \nabla_\beta \mathcal{L}_{j,\lambda}(\hat{\beta}(\lambda)) = 0.
\end{align}
Under standard regularity conditions (local convexity and invertible Hessian), the implicit function theorem guarantees that $\hat{\beta}(\lambda)$ is differentiable with respect to $\lambda$.
Differentiating the optimality condition yields
\begin{align}
\nabla_{\beta\beta} \mathcal{L}_{j,\lambda}(\hat{\beta})
\frac{d\hat{\beta}}{d\lambda}
+
\nabla_{\beta\lambda} \mathcal{L}_{j,\lambda}(\hat{\beta})
= 0.
\end{align}
Rearranging, we have
\begin{align}
\frac{d\hat{\beta}}{d\lambda}
= -\nabla_{\beta\beta} \mathcal{L}_{j,\lambda}(\hat{\beta})^{-1} \nabla_{\beta\lambda} \mathcal{L}_{j,\lambda}(\hat{\beta}).
\end{align}

Since the outer objective function 
$\mathcal{A}_j$
depends on $\hat{\beta}(\lambda)$, we compute its total derivative
\begin{align}
\nabla_\lambda
\mathcal{A}_j(\lambda)
= \nabla_{\lambda} \tilde{\mathcal{A}}_j(\hat{\beta},\lambda) + \left(\nabla_{\beta} \tilde{\mathcal{A}}_j(\hat{\beta},\lambda) \right)^\top \frac{d\hat{\beta}}{d\lambda}.    
\end{align}
And by substituting the IFT result we obtain the gradient of the outer objective
\begin{align}
\nabla_\lambda
\mathcal{A}_j(\lambda)
= 
  \nabla_{\lambda} \tilde{\mathcal{A}}_j(\hat{\beta},\lambda) - \left(\nabla_{\beta} \tilde{\mathcal{A}}_j(\hat{\beta},\lambda) \right)^\top
  \nabla_{\beta\beta} \mathcal{L}_{j,\lambda}(\hat{\beta})^{-1} \nabla_{\beta\lambda} \mathcal{L}_{j,\lambda}(\hat{\beta}).
\end{align}

\subsection{Computational considerations}
\label{subsec: computational considerations}

\begin{table}[h]
\centering
\setlength{\tabcolsep}{10pt}   %
\renewcommand{\arraystretch}{1.3}  %
\setlength{\extrarowheight}{6pt}
\begin{tabular}{|c|c|l|l|}
\hline
\textbf{Name} & \textbf{Order} & \textbf{Mathematical Expression} & \textbf{Description} \\
\hline
$\mathcal{L}_j$ & 0 & $ \mathcal{L}_j(\beta) $ & Negative log-likelihood \\
    & 1 & $ \nabla_\beta \mathcal{L}_j(\beta) $ & Gradient \\
    & 2 & $ \nabla_{\beta\beta} \mathcal{L}_j(\beta) $ & Hessian, observed Fisher information \\
\hline 
$\mathcal{L}_{j,\lambda}$ & 0 & $\mathcal{L}_{j,\lambda}(\beta, \lambda) = \mathcal{L}_{j}(\beta)+\lambda\beta^\top D^\top D \beta $ & Penalised loss as inner objective \\
    & 1 & $\nabla_\nu \mathcal{L}_{j,\lambda} (\nu) $, where $\nu = (\beta, \lambda)$ & Gradient \\
    & 2 & $ \nabla_{\nu\nu} \mathcal{L}_{j,\lambda}(\nu) $ & Hessian of penalised loss \\
\hline
$\tilde{\mathcal{A}}_j$ & 0 & $\tilde{\mathcal{A}}_j(\nu) = \mathcal{L}_j(\beta) + \texttt{tr}(\nabla_{\beta\beta} \mathcal{L}_{j,\lambda}^{-1} \nabla_{\beta\beta} \mathcal{L}_j) $ & Likelihood loss with AIC-type criterion \\
    & 1 & $ \nabla_\nu \tilde{\mathcal{A}}_j(\nu) = \nabla_\nu \mathcal{L}_j + \nabla_\nu \texttt{tr}(\nabla_{\beta\beta} \mathcal{L}_{j,\lambda}^{-1} \nabla_{\beta\beta} \mathcal{L}_j) $ & Gradient \\
\hline
$\mathcal{A}_j$ & 0 & $ \mathcal{A}_j(\lambda) = \tilde{\mathcal{A}}_j(\hat{\beta}(\lambda), \lambda)$; $\hat{\beta}(\lambda)=\arg\min_\beta \mathcal{L}_{j,\lambda}(\beta,\lambda)$ & Outer objective \\
    & 1 & $ \nabla_\lambda \mathcal{A}_j = \nabla_\lambda \tilde{\mathcal{A}}_j - \nabla_\beta \tilde{\mathcal{A}}_j^\top \nabla_{\beta\beta} \mathcal{L}_{j,\lambda}^{-1} \nabla_{\beta\lambda} \mathcal{L}_{j,\lambda} $ & Gradient of outer objective using IFT \\
\hline
\end{tabular}
\caption{Summary of objectives and their derivatives. Mostly only order 0 functions needs to be implemented while order 1 and 2 is taken care of by AD. The exception is order 1 of $\mathcal{A}_j$ in which an explicit implementation is created using the derivation in Section~\ref{subsec:gradient based adaptation}.}
\label{table:derivatives and ad}
\end{table}

To conclude, we summarise several computational aspects relevant for implementation.

All required derivatives are first- and second-order derivatives of the penalised objective $\mathcal{L}_{j,\lambda}$ and the effective degrees of freedom. These quantities are obtained via automatic differentiation \citep{griewank2008evaluating} using JAX \citep{jax2018github}.
Evaluation of the outer gradient requires only linear solves involving the Hessian $\nabla_{\beta\beta} \mathcal{L}_{j,\lambda}$, which is already computed in the inner optimisation.
Table \ref{table:derivatives and ad} provides an overview of necessary computation.

The information penalty in the outer objective (Equation~\eqref{eq:outer objective}) is modular. Here we employ the AIC, but alternative criteria such as the corrected AIC (AICc) \citep{sugiura1978further, hurvich1989regression} or the Bayesian Information Criterion (BIC) \citep{schwarz1978estimating} may be used without modification of the framework. The derivations in Section~\ref{subsec:gradient based adaptation} and Table~\ref{table:derivatives and ad} remain unchanged, as the gradient-based adaptation applies to any information criterion expressible as a function of the penalised likelihood and the effective degrees of freedom.

For coefficients associated with non-monotone components, the inner problem in Equation~\eqref{eq:inner objective} admits a closed-form solution conditional on the monotone coefficients. The derivation is provided in Appendix~ \ref{appendix:separable_optimization}. This substantially reduces the dimensionality of the inner optimisation and improves numerical stability, which is particularly important for accurate evaluation of the effective degrees of freedom in Equation~\eqref{eq:effective degrees of freedom}.

For Bayesian inference, accurate tail behaviour is essential. We therefore modify the P-spline basis to enforce linear extension beyond the boundary knots. This guarantees well-behaved extrapolation of the triangular map in the tails. Details are given in Appendix~ \ref{appendix:p-splines}.

Finally, the number of knots $K$ directly influences computational cost. Classical spline theory suggests $K=n$ for optimal approximation error of order $O(n^{-4})$ \citep{de_boor_practical_2001}. However, the dominant statistical estimation error remains $O(n^{-1/2})$\citep{wood2020inference}. Balancing approximation and estimation error yields $K=O(n^{1/5})$ \citep{claeskens2009asymptotic}. To avoid overly restricting model capacity while maintaining computational tractability, we fix $K=n^{1/3}$ (placed at empirical quantiles) and control effective complexity through penalisation.

\section{Applications and examples}
\label{sec:examples}

We demonstrate the continuous complexity adaptation algorithm across three examples of increasing realism and practical relevance.
The first is a controlled bivariate non-linear example designed to illustrate the behaviour of the inner and outer objectives.
The second considers the Lorenz-63 system \citep{Lorenz1963DeterministicFlow}, where we compare against previously tuned triangular transport maps.
The third applies the method to a distributed groundwater flow model, illustrating performance in a higher-dimensional, application driven setting.

\subsection{A bivariate nonlinear example}
\begin{figure}
    \centering
    \includegraphics[width=\linewidth]{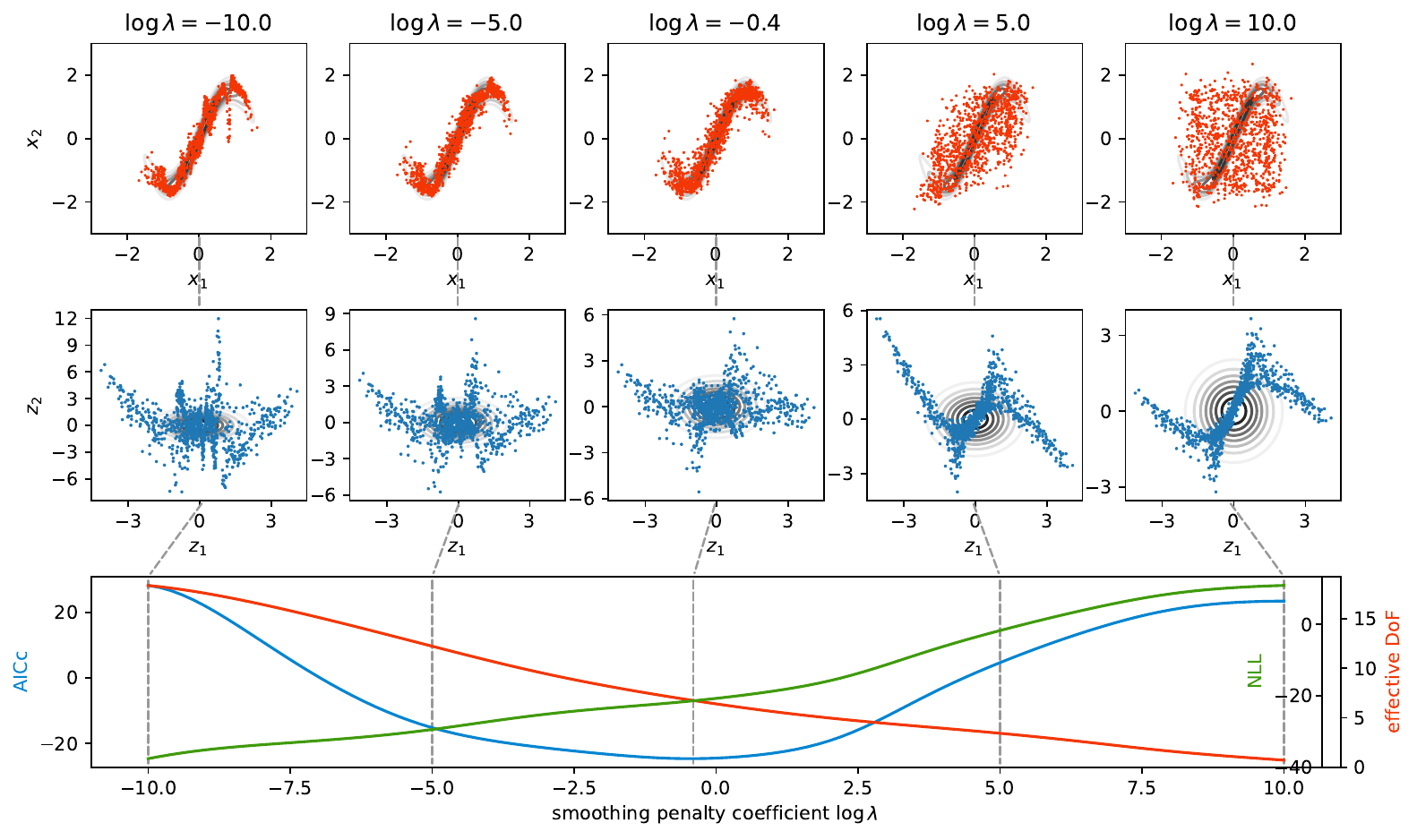}
    \caption{The effect of the smoothing penalty on the pullback $S^\#\eta$ (top row), the pushforward $S_\#\pi$ (middle row), and the AICc (bottom row). The target distribution $\pi$ is a bivariate wavy distribution. We keep smoothing penalties for the monotone parts of the map component functions $S_1$ and $S_2$ fixed at $\log\lambda=10$ and only adjust the smoothing penalty of the nonmonotone part of $S_2$. The problem is defined artificially difficult with $n=30$ ensemble size and $K=50$ knots to better highlight the influence of the smoothing penalty. For low smoothing penalties, the map tends to overfit, yielding very noisy pushforward and pullback. For high smoothing penalty, the map becomes too simple, only providing very rudimentary transformation. In-between the AIC shows an optimal trade-off between the negative log-likelihood and the effective degrees of freedom.}
    \label{fig:bivariate wavy problem}
\end{figure}

We first illustrate the proposed complexity adaptation on a synthetic bivariate target distribution. The target $\pi$ is chosen to exhibit nonlinear, oscillatory structure, creating a setting where insufficient flexibility underfits the ensemble while excessive flexibility in $S$ leads to overfitting. We examine whether the outer objective in Equation~\eqref{eq:outer objective} strikes an appropriate balance.

For each value of the smoothing parameter $\lambda$, the P-spline coefficients are obtained by solving the inner problem in Equation~\eqref{eq:inner objective}. That is, for fixed $\lambda$, the penalised negative log-likelihood is minimised to yield $\hat{\beta}(\lambda)$.

In this example, rather than optimising the outer objective, we profile it over a range of smoothing parameters. This allows us to examine how it varies with complexity and to verify that AICc identifies a meaningful trade-off between goodness-of-fit and effective degrees of freedom.

To isolate the effect of complexity control, we fix the smoothing penalties for the monotone parts of $S_1$ and $S_2$ at $\log \lambda=10$, and vary only the $\log \lambda$ penalty governing the nonmonotone part of $S_2$ within $[-10, 10]$. The problem is deliberately made data-scarce and flexible, with ensemble size $n=30$ and $K=50$ knots, so that regularisation plays a decisive role.

Figure~\ref{fig:bivariate wavy problem} shows, for five different smoothing parameters, samples from the pullback density $S^{\#}\eta$ (top row) overlaying the target, samples from the pushforward $S_{\#}\pi$ (middle row) overlaying the reference, and the profiled negative log-likelihood, effective degrees of freedom, and outer objective (bottom row).

For small smoothing penalties (left), the inner problem yields highly flexible maps that fit the ensemble excessively closely. While the negative log-likelihood is small, the effective degrees of freedom, measured by Equation~\eqref{eq:effective degrees of freedom}, are large. The outer objective penalises this complexity, yielding a large value.
The resulting pushforward deviates from the Gaussian reference away from the ensemble support, and the pullback density exhibits spurious oscillations.

For large smoothing penalties (right), the inner solution is overly constrained. The effective degrees of freedom are small, but the negative log-likelihood increases substantially. The map fails to capture the nonlinear structure of the target, leaving visible non-Gaussian structure in the pushforward.

Between these extremes, the outer objective exhibits a clear minimum (middle). At this value of $\lambda$, the map achieves a more appropriate balance between fit and smoothness: the pushforward is closest to Gaussian, and the pullback provides a stable approximation of the target. With $n=30$, the ensemble is too small to capture the nonlinear structure without overfitting to sampling noise. This confirms that the outer objective provides a sensible continuous measure of complexity.

\subsection{Lorenz-63}

In this experiment, we replicate the Ensemble Transport Filter (EnTF) simulations for the Lorenz-63 model \citep{Lorenz1963DeterministicFlow} from \citet{ramgraber2023ensemble}. In their prior study, \citet{ramgraber2023ensemble} manually regularised EnTFs of various discrete levels of complexity. In this study, we compare our automatically-adapted P-Spline EnTFs to their results.

\subsubsection{Model setup}

The Lorenz-63 system has three scalar states $\boldsymbol{x}(t) = (x^a(t),x^b(t),x^c(t))$ \citep{Lorenz1963DeterministicFlow} that evolve in time according to
\begin{equation}
    \frac{d x^{a}}{d t} = \sigma(x^{b} - x^{a}), \quad \frac{d x^{b}}{d t} = x^{a}(\rho - x^{c})-x^{b}, \quad 
    \frac{d x^{c}}{d t} = x^{a}x^{b} - \beta x^{c}.
    \label{eq:L63_scalar}
\end{equation}
The model parameters $\sigma$, $\beta$, and $\rho$ define the nature of the dynamics. Common choices are $\sigma=10$, $\beta=\frac{8}{3}$, and $\rho=28$, for which the model displays chaotic dynamics. We select these values for our simulations. We integrate the model forward in time with a fourth-order Runge-Kutta scheme, with integration step size $\Delta t = 0.05$.

\subsubsection{Data assimilation setup}

Observations arrive every $\Delta t = 0.1$ time units. We assume a perfect model and ignore forecast error. We draw the initial prior state from a trivariate standard normal distribution, then spin up the prior ensemble for 250 steps to spread it over the Lorenz attractor. We assimilate independent observations of all three states at each step for $T=1000$ steps, with Gaussian observation error standard deviation $\sigma_{\text{obs}}=0.25$.

We follow the assimilation scheme of \citet{ramgraber2023ensemble}. At each step, we assimilate the observation of each state variable separately, exploiting conditional independence between an observation and the unobserved states. For example, to assimilate $y^a$ (an observation of $x^a$), we use the triangular map
\begin{equation}
    \boldsymbol{S}\left(y^a,x^a,x^b,x^c\right) = \begin{bmatrix*}[l]
        S_1(y^a) \\
        S_2(y^a,x^a) \\
        S_3(\quad\;\,x^a,x^b) \\
        S_4(\quad\;\,x^a,x^b,x^c) \\
    \end{bmatrix*},
\end{equation}
where $y^a \ci x^b,x^c \mid x^a$. As a result, $S_3$ and $S_4$ do not depend on $y^a$. We obtain analogous sparse maps for assimilating $y^b$ and $y^c$ by permuting the state vector. This approach performs three low-dimensional updates per step instead of one higher-dimensional update, while leveraging conditional independence and enforcing independent observation errors by construction.

\subsubsection{Results}

\begin{figure}
    \centering
    \includegraphics[width=\linewidth]{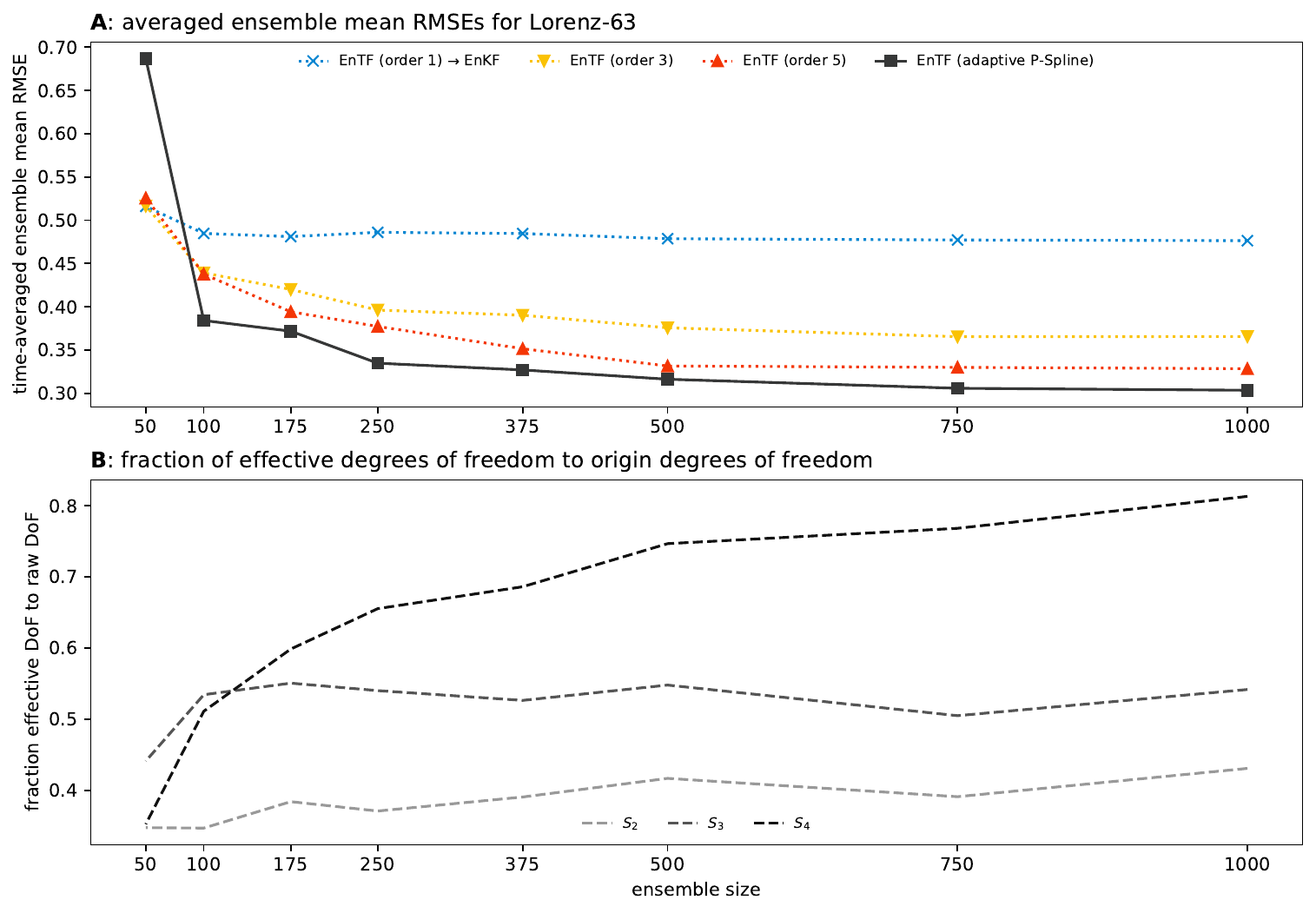}
    \caption{(A) Time-averaged ensemble mean RMSEs, averaged across ten random seeds, for EnTFs based on Hermite functions of varying maximum polynomial order \citep[blue, yellow, red; see ][]{ramgraber2023ensemble} as well as the adaptive P-spline algorithm proposed in this study. Note that the linear (order 1) EnTF corresponds to an Ensemble Kalman Filter (EnKF). (B) Averaged fraction of effective degrees of freedom to raw degrees of freedom of the adaptive P-spline EnTF for different ensemble sizes.}
    \label{fig:results_L63}
\end{figure}

Figure~\ref{fig:results_L63}A displays the ensemble mean RMSEs for EnTFs of different ensemble sizes, averaged across ten different random seeds each. We observe two important trends:
\begin{enumerate}
    \item \textbf{RMSE decreases with ensemble size}, as larger ensemble sizes reduce the sampling error and thus improve the qualtiy of the data assimilation update.
    \item \textbf{RMSE decreases with nonlinearity}, as more nonlinear maps are better capable of capturing non-Gaussian features of the distributions arising during filtering.
\end{enumerate}

In the results of \citet{ramgraber2023ensemble}, we observe different maximum polynomial order maps (order 1 to order 5), where the map's nonlinearity and complexity increases with order. This can reduce the filtering RMSE if the ensemble size is sufficiently large. For smaller ensemble sizes, the performance of the maps becomes more similar. Note, however, that the prior study used manual regularisation to avoid overfitting, and inflation to improve stability in low-ensemble size regimes. By contrast, the P-Spline EnTF results outperform the results of \citet{ramgraber2023ensemble} for all ensemble sizes except $n=50$ without the need to manually adjust either the degree of map complexity, nor the need to manually regularise the optimization. For $n=50$, 9 out of 10 random seeds also outperformed all three manual EnTFs (avg. RMSE of $0.49$ without random seed 8), but one filter diverged, increasing the average RMSE - likely a consequence of the absence of inflation in our simulations.

Figure~\ref{fig:results_L63}B shows the average fraction between the effective degrees of freedom (DoF), which is adjusted for the optimized $\lambda$, over the raw degrees of freedom, i.e., the raw number of basis functions. Results are shown across all ensemble sizes for the three map component functions $S_2$ to $S_4$ learned for the conditional inversion. We observe that the adaptive P-Spline EnTF favours simpler maps (lower fractions) for smaller ensemble sizes, and uses more of the available complexity as ensemble size grows, particularly for $S_4$. Also observe that $S_2$ remains the simplest map throughout, which makes sense considering that our observation model is linear and $S_2$ only depends on the observed state and the observation predictions. Note that both these effects occur in addition to the basis function scaling with ensemble size, which sets the number of P-spline knots to $K=n^{1/3}$.

\subsection{Groundwater model and Darcy flow}

\subsubsection{Model setup}

We demonstrate the P-Spline Ensemble Transport Smoother (P-Spline EnTS) for non-Gaussian parameter estimation in a distributed groundwater model. The experiment considers steady-state two-dimensional Darcy flow on a $51 \times 51$ regular grid with spacing $\Delta x = 1$~m and a layer thickness of $10$~m.

We impose a fixed hydraulic head of $5$~m along the eastern boundary and apply a uniform recharge of $10^{-8}$~m/s to every grid cell. Hydraulic head $h$~[m] is observed at six locations within the domain. We simulate flow with MODFLOW 6 \citep{Langevin2026MODFLOW6} using its Python interface, FloPy \citep{Bakker2026FloPy}.

\subsubsection{Data assimilation setup}

We assign a bimodal prior to the hydraulic log-conductivity, $\log_{10}K$~[$\log_{10}$ m/s]. To generate prior samples, we use the Python package \texttt{gstools} \citep{sebastian_muller_2025_15296281}. We first define a two-dimensional Gaussian random field with correlation length 10 [cell units] and draw realizations of this field. We then transform each realization to a U-quadratic distribution and rescale it to the interval $-7 \leq \log_{10}K \leq -5$.

Observation errors are independent and Gaussian with standard deviation $\sigma_{\text{obs}} = 0.01$~m. We designate one prior realization as the synthetic truth, run the groundwater model, and generate observations from its results. The remaining $n=100$ realizations form the prior ensemble.

We compare two data assimilation methods: the ensemble Kalman smoother (EnKS) \citep{evensen2000ensemble} and an adaptive P-Spline EnTS. In the P-Spline EnTS, the hydraulic heads at the six observation locations define the upper block of the triangular transport map. The lower block structure is determined by the Schäfer–Cholesky algorithm \citep{doi:10.1137/20M1336254}. The algorithm is seeded with the $\log_{10}K$ variables at the observation locations and uses a neighborhood radius of $\rho = 2$ \citep[see ][]{doi:10.1137/20M1336254}.

\subsubsection{Results}

\begin{figure}
    \centering
    \includegraphics[width=\linewidth]{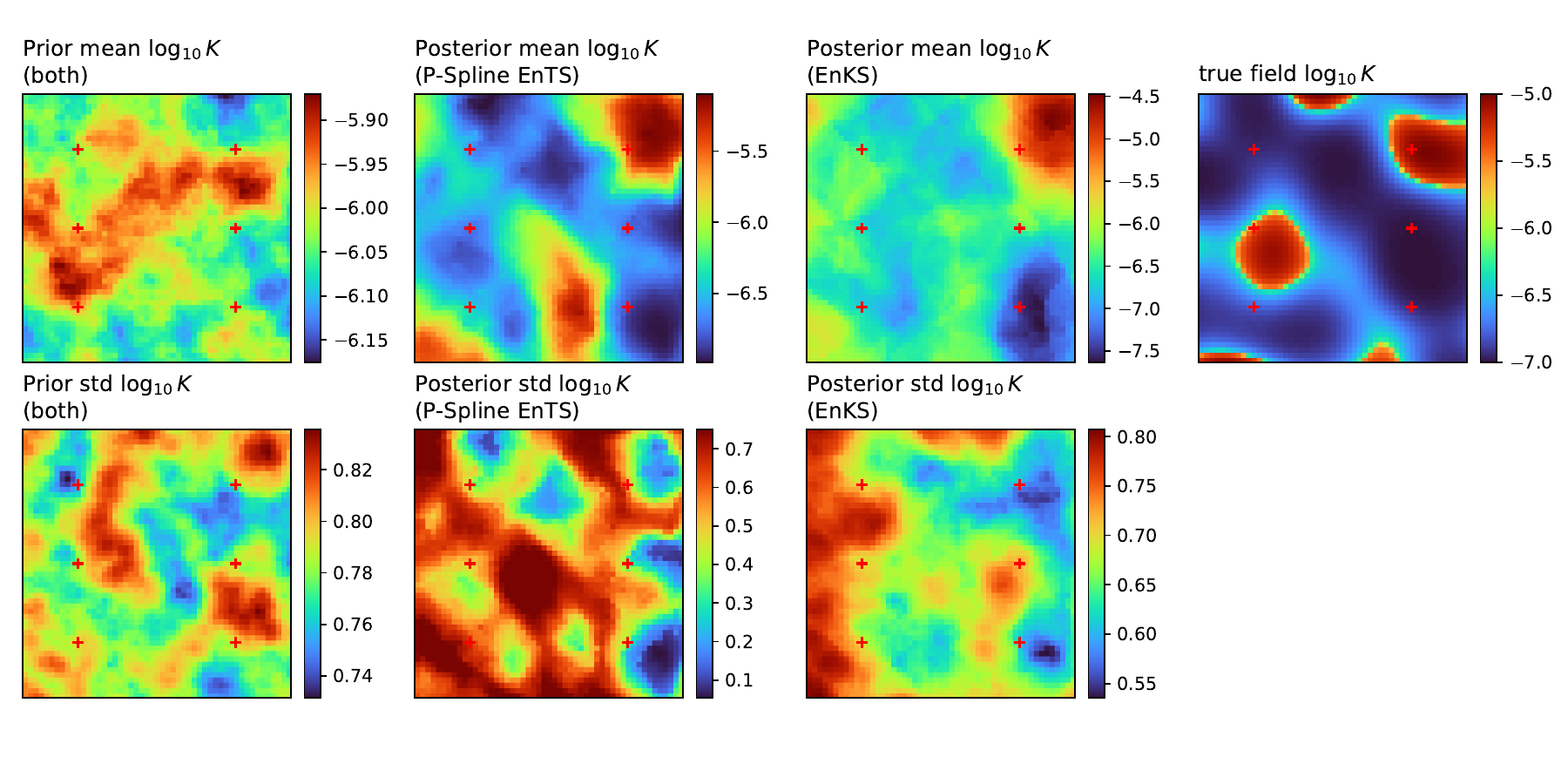}
    \caption{Prior (first column) and posterior (second and third column) mean (first row) and marginal standard deviations (second row) for the log-conductivities obtained with both methods. The synthetic truth's log hydraulic conductivity field is illustrated in the top right.}
    \label{fig:Darcy_conductivities}
\end{figure}

Figure~\ref{fig:Darcy_conductivities} shows the prior and posterior means and marginal standard deviations of the log-conductivity fields for both methods. Both posterior means reproduce the hydraulic conductivity pattern along the eastern boundary, the most sensitive region of the domain. However, the EnKS mean field for $\log_{10}K$ exceeds the prior conductivity bounds, whereas the P-Spline EnTS mean remains within them.

\begin{figure}
    \centering
    \includegraphics[width=\linewidth]{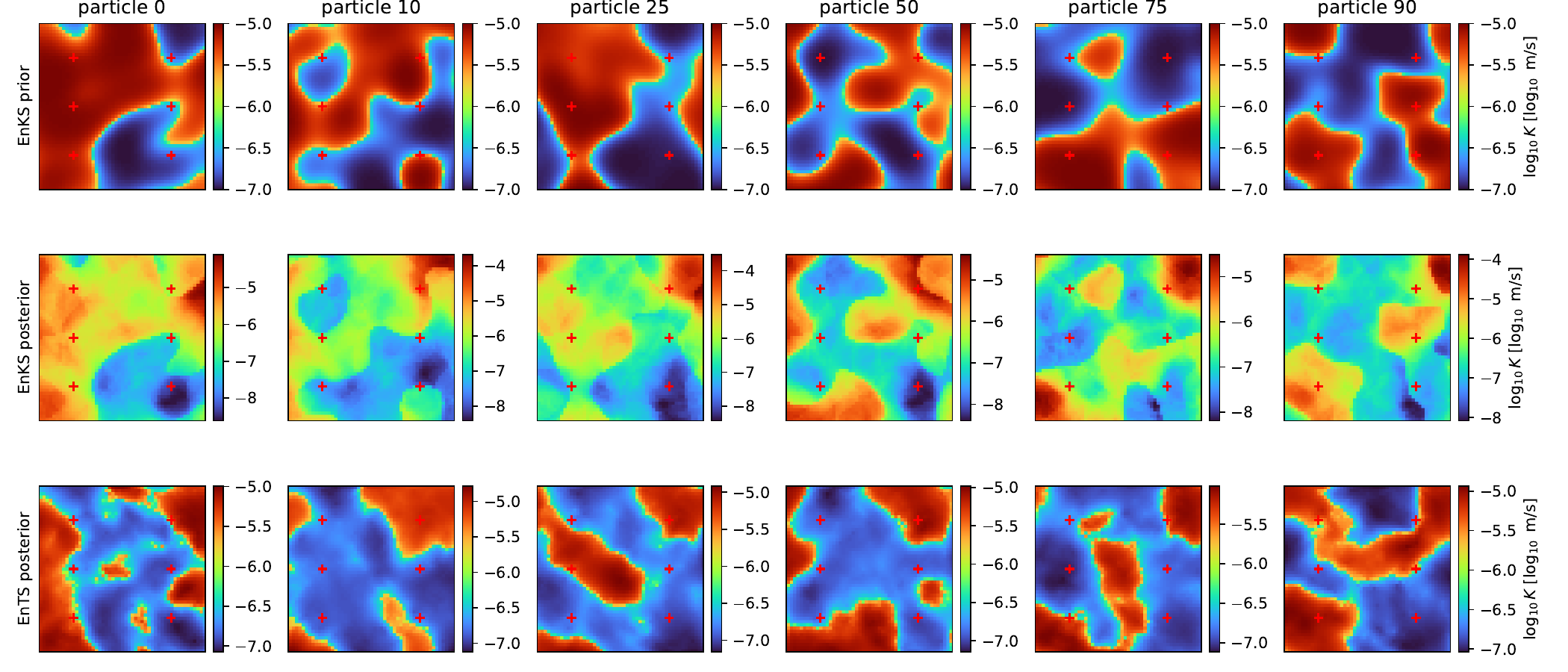}
    \caption{Prior (first row) and posterior (second row: EnKS; third row: P-Spline EnTS) log-conductivity parameter fields for six selected ensemble members. The prior log-conductivity fields are bimodal. The P-Spline EnTS preserves this feature, the EnKS does not.}
    \label{fig:posterior_conductivities_samples}
\end{figure}

Figure~\ref{fig:posterior_conductivities_samples} presents selected prior samples (top row) and posterior samples from the EnKS (middle row) and the P-Spline EnTS (bottom row). The prior log-conductivity fields are strongly bimodal. The EnKS posterior fields blur this structure and fail to preserve the distinct geological facies types. In contrast, the P-Spline EnTS adapts the facies distribution patterns while preserving bimodality, leading to better agreement with the geological prior. Both methods produce samples that violate the initial bounds $-7 \leq \log_{10}K \leq -5$~[$\log_{10}$ m/s], but the violations are substantially larger for the EnKS.

\begin{figure}
    \centering
    \includegraphics[width=\linewidth]{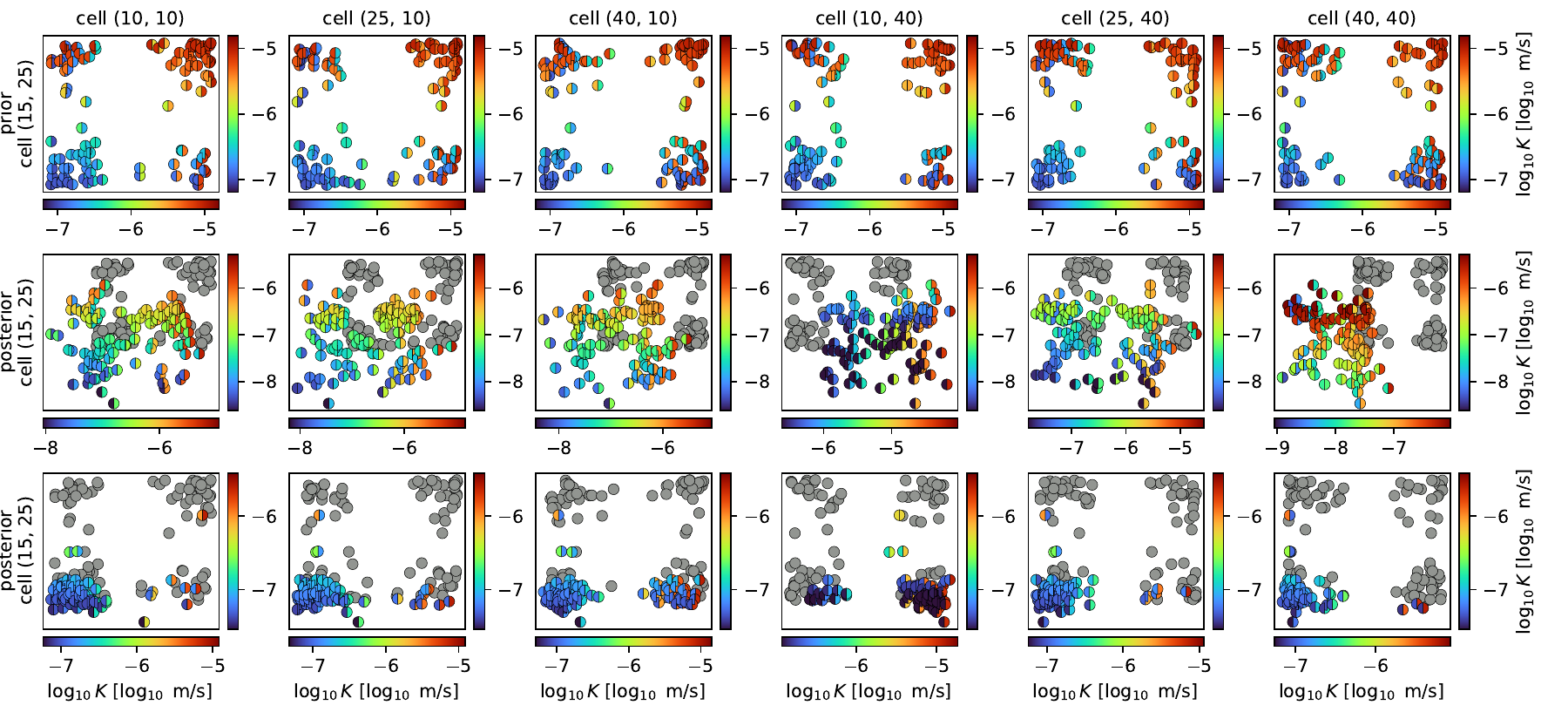}
    \caption{Bivariate marginals for the log-conductivity between grid cell $(15,25)$ and the six observation locations (columns). The first row shows the prior, the second and third row the posterior for the EnKS and P-Spline EnTS, respectively. We see that the prior ensemble is multi-modal, with each mode corresponding to different combinations of the low-conductive and high-conductive facies. The EnKS's posterior samples are mapped between the distinct modes, creating washed-out features. The P-Spline EnTS's posterior samples, by contrast, remain within the modes, preserving the bimodality.}
    \label{fig:bimodal_Darcy}
\end{figure}

Figure~\ref{fig:bimodal_Darcy} explains this behavior. The prior ensemble is multimodal (top row). Because the EnKS applies a Gaussian update, many posterior samples fall between the modes (middle row). The P-Spline EnTS, by contrast, keeps samples within the modes and produces geologically consistent realizations.

\begin{figure}
    \centering
    \includegraphics[width=\linewidth]{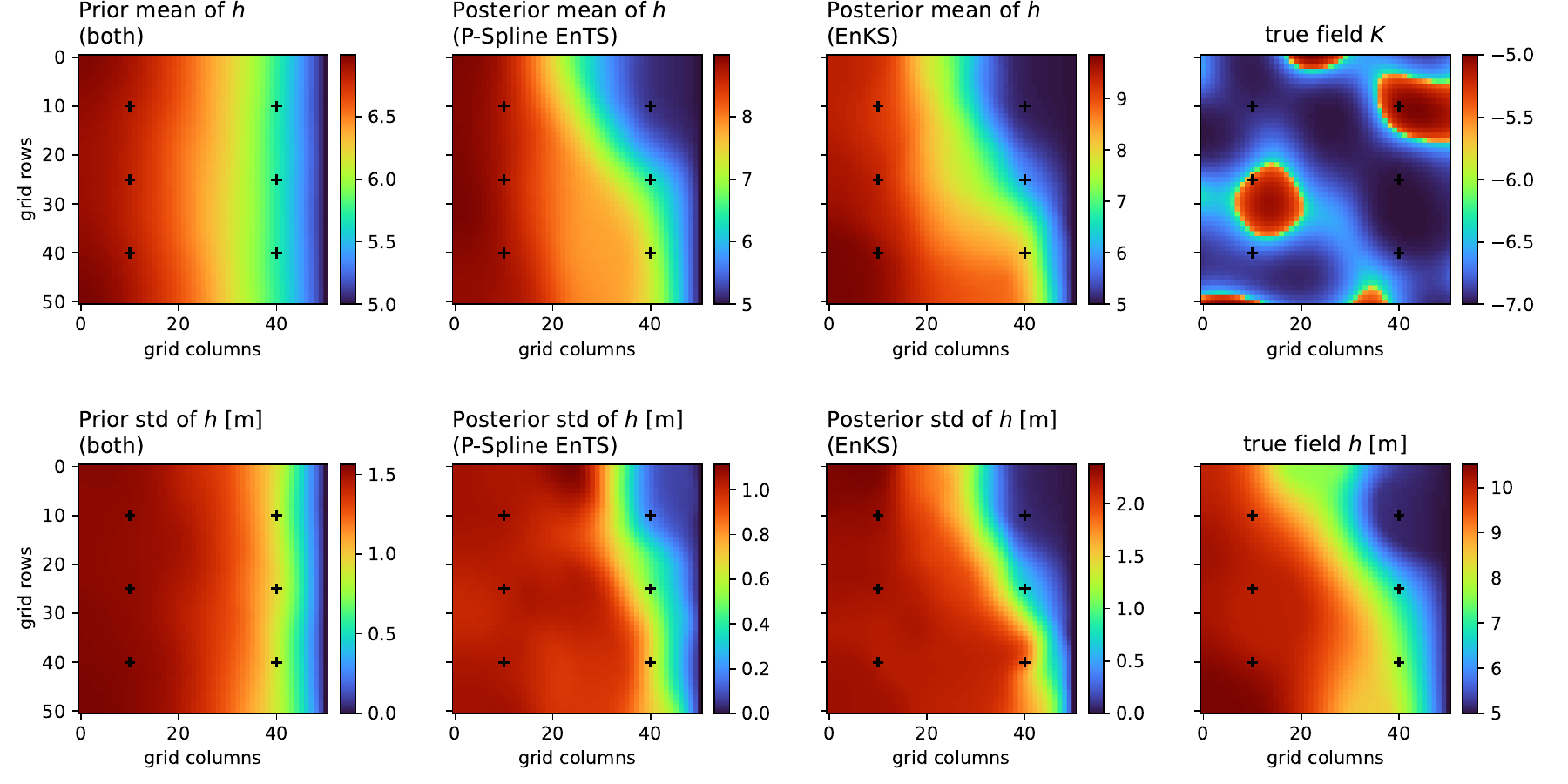}
    \caption{Mean (first row) and standard deviation (second row) of hydraulic head fields simulated from prior (first column) and posterior (second and third column) log-conductivity fields. The last column shows the true log-conductivity field (top) and the true hydraulic head field (bottom).}
    \label{fig:Darcy_heads}
\end{figure}

Finally, we re-simulate the Darcy flow with the posterior fields. Figure~\ref{fig:Darcy_heads} illustrates the prior and posterior mean and standard deviations of the resulting hydraulic head fields. We observe qualitatively that both the EnKS's and the P-Spline EnTS's mean fields replicate the true hydraulic head field well, but that standard deviations for the P-Spline EnTS are approximately half. This is reflected in the posterior simulation ensemble RMSE, which has a value of $1.23$~m for the P-Spline EnTS and a value of $1.64$~m for the EnKS. The ensemble RMSE is calculated as
\begin{equation}
    \operatorname{RMSE}\left(\{\boldsymbol{x^{(i)}}\}_{i=1}^{n}\right) = \frac{1}{n}\sum_{i=1}^{n}\left(\sqrt{\frac{1}{d}\sum_{j=1}^{d}\left(x^{(i)}_{j} - x^{\operatorname{true}}_{j}\right)^2}\right),
\end{equation}
where $\boldsymbol{x}^{(i)}=[x^{(i)}_1,\dots,x^{(i)}_d]^\intercal$ is the $i$-th ensemble member's state vector obtained from a simulation with the posterior log-conductivity fields.

\section{Conclusion}
\label{sec:conclusions}

In this manuscript, we developed a continuous complexity adaptation algorithm for nonlinear triangular transport maps. Through a smoothing penalty coefficient, we can continuously adapt the degrees of freedom of a triangular map parametrized through P-splines. Leveraging an information criterion as a measure of continuous complexity, we efficiently and automatically identify an optimal level of map complexity through gradient descent.

We demonstrated the performance of the resulting algorithm in three numerical experiments. First, we illustrated the consequences of under- and overfitting in a bivariate test distribution. Second, we demonstrated the performance of the algorithm for nonlinear filtering in a Lorenz-63 system and compared the results to a previous study. Finally, we illustrated the method's scalability for history matching in a high-dimensional, distributed groundwater model. 

Limitations include computational demand: while the adaptation strategy itself is efficient, it is based on the construction and evaluation of a full P-spline basis for each map component function $S_i$. In consequence, reductions in map complexity do not cause a reduction in computational demand. To increase efficiency, we also hypothesize that it may not be necessary to adapt the monotone part of each map component function, since the monotone reparametrisation already exerts a regularising influence.

To ensure scalability in triangular maps, it is imperative to efficiently detect and exploit conditional independence in high-dimensional systems. In this study, we leverage the sparse Cholesky algorithm by \citet{doi:10.1137/20M1336254}, which proposes an efficient ordering for spatial variables. However, in many systems, it may be necessary to leverage graph detection strategies that can also deal efficiently with variables that do not have spatial or temporal coordinates. To this end, location-agnostic strategies such as coordinate gradient boosting or LASSO seem promising directions for future research.

The algorithm developed in this study permits efficient plug-and-play nonlinear data assimilation, which automates the search for a parsimonious map parametrisation without requiring manual hyperparameter tuning. We observe that this algorithm can match and even outperform manually tuned nonlinear triangular maps. We also show that its parsimonious nonlinear updates allow the posterior samples to honor complex geological priors significantly better than linear-Gaussian alternatives.

\section{Code availability}

The code to reproduce the figures and simulations in this manuscript is available on GitHub under: \url{https://github.com/MaxRamgraber/Adaptive-P-Spline-Triangular-Measure-Transport}

\section{Acknowledgements} \label{acknowledgements}

The research of MR leading to these results has received funding from the Dutch Research Council under the Veni grant VI.Veni.232.140. The research of BL leading to these results is supported in part by the Adaptive AI project funded by Equinor through Akademiaavtalen. MR and BL acknowledge support through Equinor's DATeS project. MR and BL want to thank the UQ group of MIT for many insightful discussions related to the research in this manuscript.

\appendix
\clearpage

\newcommand{\x}{\boldsymbol{x}}

\section{Optimizing linear separable maps}\label{appendix:separable_optimization}

Here, we reproduce and adapt the derivation in the Appendix of \citet{ramgraber_friendly_2025} with the parametrization proposed in this manuscript. Choosing a \textit{linear separable} map parameterization permits more efficient map optimization. First, for the purpose of this material we denote gradient of some scalar-valued function $r(\mathbf{z})$ with respect to $\mathbf{z}$ evaluated at $\mathbf{z}^*$ as $\nabla_\mathbf{z} r\big|_{\mathbf{z}^*}$ and the derivative of a univariate function $s(z)$ with respect to scalar $z$ evaluated at $z^*$ as $\partial_z s\big|_{z^*}$. Then, recall for a separable map that is \textit{linear in the coefficients}, each map component function $S_{k}$ is defined as:

\begin{equation}
\begin{aligned}
    S_{k}\left(x_{1},\dots,x_{k-1},x_{k}\right) &= g(\x_{1:k-1}) + f(x_{k})\\
    &= \beta_{k,1}^{\text{non}}B_{k,1}^{\text{non}}(\x_{1:k-1}) + \dots + \beta_{k,m}^{\text{non}}B_{k,m}^{\text{non}}(\x_{1:k-1}) + \beta_{k,1}^{\text{mon}}B_{k,1}^{\text{mon}}(x_{k}) + \dots + \beta_{k,n}^{\text{mon}}B_{k,n}^{\text{mon}}(x_{k}) \\
    &= \boldsymbol{B}_{k}^{\text{non}}(\x_{1:k-1}) \boldsymbol{\beta}_{k}^{\text{non}} + \boldsymbol{B}_{k}^{\text{mon}}(x_{k})\boldsymbol{\beta}_{k}^{\text{mon}}
\end{aligned}
\label{apeq:separable_map_discrete}
\end{equation}

where $B_{k,j}^{\text{non}}:\mathbb{R}^{k-1}\to\mathbb{R}$ and $B_{k,j}^{\text{mon}}:\mathbb{R}\to\mathbb{R}$ are the $j$-th basis functions of map component $S_{k}$, associated with the nonmonotone and monotone terms, respectively. Then, $\boldsymbol{B}_{k}^{\text{non}}:\mathbb{R}^{k-1}\to \mathbb{R}^{1\times m}$ and $\boldsymbol{B}_{k}^{\text{mon}}:\mathbb{R}\to \mathbb{R}^{1\times n}$ are vectors of basis function evaluations for $g$ and $f$, and $\boldsymbol{\beta}_{k}^{\text{non}}\in\mathbb{R}^{ m \times 1}$ and $\boldsymbol{\beta}_{k}^{\text{mon}}\in\mathbb{R}^{ n \times 1}$ are the corresponding column vectors of coefficients. Therefore, the expressions $\boldsymbol{B}_k^{\text{non}}(\x_{1:k-1})\boldsymbol{\beta}^{\text{non}}_k$ and $\boldsymbol{B}_k^{\text{mon}}(x_k)\boldsymbol{\beta}_k^{\text{mon}}$ are both inner product functions of $\x$, i.e. scalar-valued. For \textit{maps from samples} \citep[see ][]{ramgraber_friendly_2025}, we consider samples $\boldsymbol{\mathsf{X}}^{1},\ldots,\boldsymbol{\mathsf{X}}^{N}\sim\pi$ and recall the following optimization objective for $S_{k}$:

\begin{equation}
    \mathcal{J}_{k}(S_{k}) = \sum_{i=1}^{N}\left(\frac{1}{2}S_{k}(\boldsymbol{\mathsf{X}}^i)^{2} - \log\partial_{x_k}S_k\big|_{\boldsymbol{\mathsf{X}}^i} \right)
\end{equation}

Plugging in Equation~\ref{apeq:separable_map_discrete}, we obtain
\begin{equation}
\begin{aligned}
    \mathcal{J}_k(S_k) &= \sum_{i=1}^N \frac{1}{2}\left[\boldsymbol{B}_k^{\text{non}}(\boldsymbol{\mathsf{X}}_{1:k-1}i)\boldsymbol{\beta}_k^{\text{non}} + \boldsymbol{B}_k^{\text{mon}}(\boldsymbol{\mathsf{X}}_{k}^{i})\boldsymbol{\beta}_k^{\text{mon}} \right]^2 - \log\partial_{x_k}\left[\boldsymbol{B}_k^{\text{mon}}(\boldsymbol{\mathsf{X}}_{1:k-1}i)\boldsymbol{\beta}_k^{\text{non}} + \boldsymbol{B}_k^{\text{mon}}(\boldsymbol{\mathsf{X}}_{k})\boldsymbol{\beta}_k^{\text{mon}}\right]_{\boldsymbol{\mathsf{X}}_{k}^{i}} \\
    &= \sum_{i=1}^N \frac{1}{2}\left[\boldsymbol{B}_k^{\text{non}}(\boldsymbol{\mathsf{X}}_{1:k-1}i)\boldsymbol{\beta}_k^{\text{non}} + \boldsymbol{B}_k^{\text{mon}}(\boldsymbol{\mathsf{X}}_{k}^{i})\boldsymbol{\beta}_k^{\text{mon}} \right]^2 - \log\partial_{x_k}\boldsymbol{B}_k^{\text{mon}}\big|_{\boldsymbol{\mathsf{X}}_{k}^{i}}\boldsymbol{\beta}_k^{\text{mon}}.
\end{aligned}\label{apeq:plugin}
\end{equation}

Likewise, we can formulate a penalized objective:

\begin{equation}
    \mathcal{J}_k(S_k)^{\text{pen.}} = \sum_{i=1}^N \frac{1}{2}\left[\boldsymbol{B}_k^{\text{non}}(\boldsymbol{\mathsf{X}}_{1:k-1}i)\boldsymbol{\beta}_k^{\text{non}} + \boldsymbol{B}_k^{\text{mon}}(\boldsymbol{\mathsf{X}}_{k}^{i})\boldsymbol{\beta}_k^{\text{mon}} \right]^2 - \log\partial_{x_k}\boldsymbol{B}_k^{\text{mon}}\big|_{\boldsymbol{\mathsf{X}}_{k}^{i}}\boldsymbol{\beta}_k^{\text{mon}} + \frac{1}{2}{\boldsymbol{\beta}_k^{\text{non}}}^\intercal\mathbf{S}^{\text{non}}\boldsymbol{\beta}_k^{\text{non}} + \frac{1}{2}{\boldsymbol{\beta}_k^{\text{mon}}}^\intercal\mathbf{S}^{\text{mon}}\boldsymbol{\beta}_k^{\text{mon}}.\label{apeq:plugin_with_pen}
\end{equation}

Note that the samples $\boldsymbol{\mathsf{X}}^i$ are defined by the target distribution $\pi$, and are thus fixed during optimization. In consequence, the basis function evaluation vectors $\boldsymbol{B}_k^{\text{non}}$ and $\boldsymbol{B}_k^{\text{mon}}$ are also fixed for a given map parameterization, and are thus independent of the coefficients $\boldsymbol{\beta}_k^{\text{non}}$ and $\boldsymbol{\beta}_k^{\text{mon}}$. We can simplify Equation~\ref{apeq:plugin} further by absorbing the sum over the first term, and defining a new variable for the partial derivative in the second term. To this end, we form matrices $\mathbf{P}_{k}^{\text{non}}\in\mathbb{R}^{N\times m}$ and $\mathbf{P}_{k}^{\text{mon}}\in\mathbb{R}^{N\times n}$, and vectors $\mathbf{b}_{k}\in\mathbb{R}^{1\times n}$, which are each defined element-wise as
\[[\mathbf{P}_{k}^{\text{non}}]_{ij}=B^{\text{non}}_{k,j}(\boldsymbol{\mathsf{X}}_{1:k-1}i), \quad [\mathbf{P}_{k}^{\text{mon}}]_{ij}=B^{\text{mon}}_{k,j}(\boldsymbol{\mathsf{X}}_{k}^{i}),\quad [\mathbf{b}_{k}]_j = \partial_{x_k}B^{\text{mon}}_{k,j}\big|_{\boldsymbol{\mathsf{X}}_{k}^{i}},\]
where the $ij$ entry is the $j$th indexed basis function evaluated at sample index $i$ for both matrices $\mathbf{P}_{k}^{\text{non}}$ and $\mathbf{P}_{k}^{\text{mon}}$, and the log of the derivative of $\boldsymbol{B}_k^{\text{mon}}$ with respect to $\boldsymbol{\mathsf{X}}_{k}$ is evaluated at $\boldsymbol{\mathsf{X}}_{k}^{i}$ for the vectors $\mathbf{b}_{k}$. As above, these matrices and vectors can be pre-computed. The optimization objective now simplifies further to

\[\mathcal{J}_k(S_k) = \frac{1}{2}\left\|\mathbf{P}_{k}^{\text{non}}\boldsymbol{\beta}_k^{\text{non}} + \mathbf{P}_{k}^{\text{mon}}\boldsymbol{\beta}_k^{\text{mon}}\right\|^2 - \sum_{i=1}^N \log\mathbf{b}_{k}\boldsymbol{\beta}_k^{\text{mon}}.\]

Likewise, the penalized objective simplifies to

\[\mathcal{J}_k(S_k)^{\text{pen.}} = \frac{1}{2}\left\|\mathbf{P}_{k}^{\text{non}}\boldsymbol{\beta}_k^{\text{non}} + \mathbf{P}_{k}^{\text{mon}}\boldsymbol{\beta}_k^{\text{mon}}\right\|^2 - \sum_{i=1}^N \log\mathbf{b}_{k}\boldsymbol{\beta}_k^{\text{mon}} + \frac{1}{2}{\boldsymbol{\beta}_k^{\text{non}}}^\intercal\mathbf{S}^{\text{non}}\boldsymbol{\beta}_k^{\text{non}} + \frac{1}{2}{\boldsymbol{\beta}_k^{\text{mon}}}^\intercal\mathbf{S}^{\text{mon}}\boldsymbol{\beta}_k^{\text{mon}}.\]

This function is similar to that of an objective for an interior point method and, remarkably, becomes quadratic in $\boldsymbol{\beta}_k^{\text{non}}$. At this point, we add L2 (i.e., Tikhonov) regularization on both $\boldsymbol{\beta}_k^{\text{mon}}$ and $\boldsymbol{\beta}_k^{\text{non}}$ according to the guidance in \citet{ramgraber_friendly_2025}, Section~3.
\begin{equation}\label{apeq:reg_loss}
\mathcal{J}_k(S_k;\lambda) = \frac{1}{2}\left\|\mathbf{P}_{k}^{\text{non}}\boldsymbol{\beta}_k^{\text{non}} + \mathbf{P}_{k}^{\text{mon}}\boldsymbol{\beta}_k^{\text{mon}}\right\|^2 - \sum_{i=1}^N \log\mathbf{b}_{k}\boldsymbol{\beta}_k^{\text{mon}} + \frac{\lambda}{2}(\|\boldsymbol{\beta}_k^{\text{non}}\|^2 + \|\boldsymbol{\beta}_k^{\text{mon}}\|^2)
\end{equation}
In practice, this means the optimal coefficients $\widehat{\boldsymbol{\beta}}_k^{\text{non}}$ for the nonmonotone basis function evaluations minimizing \eqref{apeq:reg_loss} must satisfy
\begin{align}
0 &\equiv \nabla_{\boldsymbol{\beta}_k^{\text{non}}} \mathcal{J}_k(S_k;\lambda) \big|_{\widehat{\boldsymbol{\beta}}_k^{\text{non}}}\nonumber\\
&= \nabla_{\boldsymbol{\beta}_k^{\text{non}}}\frac{1}{2}\left\|\mathbf{P}_{k}^{\text{non}}\boldsymbol{\beta}_k^{\text{non}} + \mathbf{P}_{k}^{\text{mon}}\boldsymbol{\beta}_k^{\text{mon}}\right\|^2\bigg|_{\widehat{\boldsymbol{\beta}}_k^{\text{non}}}+\lambda\boldsymbol{\beta}_k^{\text{non}}\nonumber\\
&= ({\mathbf{P}_{k}^{\text{non}}}^{\top}\mathbf{P}_{k}^{\text{non}}+\lambda\mathbf{I})\widehat{\boldsymbol{\beta}}_k^{\text{non}} + {\mathbf{P}_{k}^{\text{non}}}^\top\mathbf{P}_{k}^{\text{mon}}\boldsymbol{\beta}_k^{\text{mon}}.\label{apeq:nonk_loss}
\end{align}

And equivalently:

\begin{equation}
\begin{aligned}
0 &\equiv \nabla_{\boldsymbol{\beta}_k^{\text{non}}} \mathcal{J}_k(S_k;\lambda)^{\text{pen.}} \big|_{\widehat{\boldsymbol{\beta}}_k^{\text{non}}}\nonumber\\
&= ({\mathbf{P}_{k}^{\text{non}}}^{\top}\mathbf{P}_{k}^{\text{non}}+ \mathbf{S}^{\text{non}})\widehat{\boldsymbol{\beta}}_k^{\text{non}} + {\mathbf{P}_{k}^{\text{non}}}^\top\mathbf{P}_{k}^{\text{mon}}\boldsymbol{\beta}_k^{\text{mon}}.
\end{aligned}
\end{equation}

In consequence, for a given choice of coefficients parameterizing the monotone functions, $\boldsymbol{\beta}_k^{\text{mon}}$, we can find the optimal choice of $\widehat{\boldsymbol{\beta}}_k^{\text{non}}$ by solving the \textit{normal equations}; for this scenario, we assume that $m < N$, i.e. the number of samples surpasses the number of basis functions (and thus ${\mathbf{P}_{k}^{\text{non}}}^\top\mathbf{P}_{k}^{\text{non}}$ is full rank).
The normal equations are a well-studied class of problems in numerical linear algebra.
Assuming linearly independent basis functions, this affords a solution for $\widehat{\boldsymbol{\beta}}_k^{\text{non}}$ as a function of $\boldsymbol{\beta}_k^{\text{mon}}$ given as

\begin{equation}\label{apeq:cnon_hat}
\begin{aligned}
({\mathbf{P}_{k}^{\text{non}}}^\top\mathbf{P}_{k}^{\text{non}}+\lambda\mathbf{I})\widehat{\boldsymbol{\beta}}_k^{\text{non}} &= -{\mathbf{P}_{k}^{\text{non}}}^\top\mathbf{P}_{k}^{\text{mon}}\boldsymbol{\beta}_k^{\text{mon}} \\ \widehat{\boldsymbol{\beta}}_k^{\text{non}} &= -\underbrace{({\mathbf{P}_{k}^{\text{non}}}^\top\mathbf{P}_{k}^{\text{non}}+\lambda\mathbf{I})^{-1}{\mathbf{P}_{k}^{\text{non}}}^\top}_{\mathbf{M}_{k,\lambda}}\mathbf{P}_{k}^{\text{mon}}\boldsymbol{\beta}_k^{\text{mon}}\\
&\stackrel{\text{def}}{=} -\mathbf{M}_{k,\lambda}\mathbf{P}_{k}^{\text{mon}}\boldsymbol{\beta}_k^{\text{mon}}
\end{aligned}
\end{equation}

And likewise

\begin{equation}
\begin{aligned}
({\mathbf{P}_{k}^{\text{non}}}^{\top}\mathbf{P}_{k}^{\text{non}}+ \mathbf{S}^{\text{non}})\widehat{\boldsymbol{\beta}}_k^{\text{non}} &= - {\mathbf{P}_{k}^{\text{non}}}^\top\mathbf{P}_{k}^{\text{mon}}\boldsymbol{\beta}_k^{\text{mon}} \\
\widehat{\boldsymbol{\beta}}_k^{\text{non}} &= - \underbrace{\left(({\mathbf{P}_{k}^{\text{non}}}^{\top}\mathbf{P}_{k}^{\text{non}}+ \mathbf{S}^{\text{non}})\right)^{-1}{\mathbf{P}_{k}^{\text{non}}}^\top}_{\mathbf{M}_{k,\mathbf{S}^{\text{non}}}} \mathbf{P}_{k}^{\text{mon}}\boldsymbol{\beta}_k^{\text{mon}} \\
&\stackrel{\text{def}}{=} -\mathbf{M}_{k,\mathbf{S}^{\text{non}}}\mathbf{P}_{k}^{\text{mon}}\boldsymbol{\beta}_k^{\text{mon}}
\end{aligned}
\end{equation}

Following best practices from numerical linear algebra, this inversion should not be computed explicitly; rather, one should use a numerical solver for the systems induced. Then, substituting Expression~\eqref{apeq:cnon_hat} into the optimization objective, we obtain a new objective for $\boldsymbol{\beta}_k^{\text{mon}}$ (i.e., entirely independent from the nonmonotone coefficients),
\begin{align}
    \mathcal{J}_k^{\text{mon}}(\boldsymbol{\beta}_k^{\text{mon}};\lambda) &= \frac{1}{2}\|-\mathbf{P}_{k}^{\text{non}}\mathbf{M}_{k,\lambda}\mathbf{P}_{k}^{\text{mon}}\boldsymbol{\beta}_k^{\text{mon}} + \mathbf{P}_{k}^{\text{mon}}\boldsymbol{\beta}_k^{\text{mon}}\|^2 - \sum_{i=1}^N \log\mathbf{b}_{k}\boldsymbol{\beta}_k^{\text{mon}} + \frac{\lambda}{2}(\|\widehat{\boldsymbol{\beta}}_k^{\text{non}}\|^2+\|\boldsymbol{\beta}_k^{\text{mon}}\|^2)\nonumber\\
    &= \frac{1}{2}\|\underbrace{(\mathbf{I}-\mathbf{P}_{k}^{\text{non}}\mathbf{M}_{k,\lambda})\mathbf{P}_{k}^{\text{mon}}}_{\mathbf{A}_{k,\lambda}}\boldsymbol{\beta}_k^{\text{mon}}\|^2  - \sum_{i=1}^N\log\mathbf{b}_{k}\boldsymbol{\beta}_k^{\text{mon}} + \frac{\lambda}{2}(\|\underbrace{\mathbf{M}_{k,\lambda}\mathbf{P}_{k}^{\text{mon}}}_{\mathbf{D}_{k,\lambda}}\boldsymbol{\beta}_k^{\text{mon}}\|^2 + \|\boldsymbol{\beta}_k^{\text{mon}}\|^2)\nonumber\\
    &\stackrel{\text{def}}{=} \frac{1}{2}\|\mathbf{A}_{k,\lambda}\boldsymbol{\beta}_k^{\text{mon}}\|^2 - \sum_{i=1}^N\log\mathbf{b}_{k}\boldsymbol{\beta}_k^{\text{mon}} + \frac{\lambda}{2}(\|\mathbf{D}_{k,\lambda}\boldsymbol{\beta}_k^{\text{mon}}\|^2 + \|\boldsymbol{\beta}_k^{\text{mon}}\|^2),
\end{align}

Which means that for the penalized objective, we get:

\begin{equation}
\begin{aligned}
    \mathcal{J}_k(S_k)^{\text{pen.}} &= \frac{1}{2}\left\|\mathbf{P}_{k}^{\text{non}}\boldsymbol{\beta}_k^{\text{non}} + \mathbf{P}_{k}^{\text{mon}}\boldsymbol{\beta}_k^{\text{mon}}\right\|^2 - \sum_{i=1}^N \log\mathbf{b}_{k}\boldsymbol{\beta}_k^{\text{mon}} + \frac{1}{2}{\boldsymbol{\beta}_k^{\text{non}}}^\intercal\mathbf{S}^{\text{non}}\boldsymbol{\beta}_k^{\text{non}} + \frac{1}{2}{\boldsymbol{\beta}_k^{\text{mon}}}^\intercal\mathbf{S}^{\text{mon}}\boldsymbol{\beta}_k^{\text{mon}}\\
    &= \frac{1}{2}\left\|\mathbf{P}_{k}^{\text{non}}\left(-\mathbf{M}_{k,\mathbf{S}^{\text{non}}}\mathbf{P}_{k}^{\text{mon}}\boldsymbol{\beta}_k^{\text{mon}}\right) + \mathbf{P}_{k}^{\text{mon}}\boldsymbol{\beta}_k^{\text{mon}}\right\|^2 - \sum_{i=1}^N \log\mathbf{b}_{k}\boldsymbol{\beta}_k^{\text{mon}} \\&\qquad + \frac{1}{2}{(\mathbf{M}_{k,\mathbf{S}^{\text{non}}}\mathbf{P}_{k}^{\text{mon}}\boldsymbol{\beta}_k^{\text{mon}})}^\intercal\mathbf{S}^{\text{non}}\mathbf{M}_{k,\mathbf{S}^{\text{non}}}\mathbf{P}_{k}^{\text{mon}}\boldsymbol{\beta}_k^{\text{mon}} + \frac{1}{2}{\boldsymbol{\beta}_k^{\text{mon}}}^\intercal\mathbf{S}^{\text{mon}}\boldsymbol{\beta}_k^{\text{mon}}\\
    &= \frac{1}{2}\left\|\underbrace{\left(\mathbf{I} - \mathbf{P}_{k}^{\text{non}}\mathbf{M}_{k,\mathbf{S}^{\text{non}}}\right)\mathbf{P}_{k}^{\text{mon}}}_{\mathbf{A}_{k,\mathbf{S}^{\text{non}}}}\boldsymbol{\beta}_k^{\text{mon}}\right\|^2 - \sum_{i=1}^N \log\mathbf{b}_{k}\boldsymbol{\beta}_k^{\text{mon}} \\&\qquad + \frac{1}{2}\boldsymbol{\beta}_k^{\text{mon},\intercal}\mathbf{P}_{k}^{\text{mon},\intercal}\mathbf{M}_{k,\mathbf{S}^{\text{non}}}^\intercal\mathbf{S}^{\text{non}}\underbrace{\mathbf{M}_{k,\mathbf{S}^{\text{non}}}\mathbf{P}_{k}^{\text{mon}}}_{\mathbf{D}_{k,\mathbf{S}^{\text{non}}}}\boldsymbol{\beta}_k^{\text{mon}} + \frac{1}{2}{\boldsymbol{\beta}_k^{\text{mon}}}^\intercal\mathbf{S}^{\text{mon}}\boldsymbol{\beta}_k^{\text{mon}}\\
    &= \frac{1}{2}\left\|\mathbf{A}_{k,\mathbf{S}^{\text{non}}}\boldsymbol{\beta}_k^{\text{mon}}\right\|^2 - \sum_{i=1}^N \log\mathbf{b}_{k}\boldsymbol{\beta}_k^{\text{mon}} \\&\qquad + \frac{1}{2}\boldsymbol{\beta}_k^{\text{mon},\intercal}\mathbf{D}_{k,\mathbf{S}^{\text{non}}}^\intercal\mathbf{S}^{\text{non}}\mathbf{D}_{k,\mathbf{S}^{\text{non}}}\boldsymbol{\beta}_k^{\text{mon}} + \frac{1}{2}{\boldsymbol{\beta}_k^{\text{mon}}}^\intercal\mathbf{S}^{\text{mon}}\boldsymbol{\beta}_k^{\text{mon}}\\
\end{aligned}    
\end{equation}

where $\mathbf{A}_{k,\mathbf{S}^{\text{non}}}$, $\mathbf{D}_{k,\mathbf{S}^{\text{non}}}$ and $\mathbf{b}_{k}$ can be precomputed prior to the optimization routine via evaluation of the basis functions. In short:

\begin{equation}
\begin{aligned}
    \mathcal{J}_k(S_k)^{\text{pen.}} 
    &= \frac{1}{2}\left\|\mathbf{A}_{k,\mathbf{S}^{\text{non}}}\boldsymbol{\beta}_k^{\text{mon}}\right\|^2 - \sum_{i=1}^N \log\mathbf{b}_{k}\boldsymbol{\beta}_k^{\text{mon}} + \frac{1}{2}\boldsymbol{\beta}_k^{\text{mon},\intercal}\mathbf{D}_{k,\mathbf{S}^{\text{non}}}^\intercal\mathbf{S}^{\text{non}}\mathbf{D}_{k,\mathbf{S}^{\text{non}}}\boldsymbol{\beta}_k^{\text{mon}} + \frac{1}{2}{\boldsymbol{\beta}_k^{\text{mon}}}^\intercal\mathbf{S}^{\text{mon}}\boldsymbol{\beta}_k^{\text{mon}}\\
    &= \frac{1}{2}\left\|\mathbf{A}_{k,\mathbf{S}^{\text{non}}}\boldsymbol{\beta}_k^{\text{mon}}\right\|^2 - \sum_{i=1}^N \log\mathbf{b}_{k}\boldsymbol{\beta}_k^{\text{mon}} + \frac{1}{2}\boldsymbol{\beta}_k^{\text{mon},\intercal}\left(\mathbf{D}_{k,\mathbf{S}^{\text{non}}}^\intercal\mathbf{S}^{\text{non}}\mathbf{D}_{k,\mathbf{S}^{\text{non}}} + \mathbf{S}^{\text{mon}}\right)\boldsymbol{\beta}_k^{\text{mon}} \\
\end{aligned}
\end{equation}

Remarkably, this method translates the original loss function into a very simple constrained convex optimization problem
\begin{equation}\label{apeq:overdet_soln}
\boxed{
\widehat{\boldsymbol{\beta}}_k^{\text{mon}} = \argmin_{\boldsymbol{\beta}_k^{\text{mon}}\geq \mathbf{0}} \frac{1}{2}\|\mathbf{A}_{k,\lambda}\boldsymbol{\beta}_k^{\text{mon}}\|^2 - \sum_{i=1}^N \log\mathbf{b}_{k}\boldsymbol{\beta}_k^{\text{mon}} + \frac{\lambda}{2}(\|\mathbf{D}_{k,\lambda}\boldsymbol{\beta}_k^{\text{mon}}\|^2 + \|\boldsymbol{\beta}_k^{\text{mon}}\|^2),\quad \widehat{\boldsymbol{\beta}}_k^{\text{non}} = -\mathbf{M}_{k,\lambda}\mathbf{P}_{k}^{\text{mon}}\widehat{\boldsymbol{\beta}}_k^{\text{mon}}.}
\end{equation}
and equivalently
\begin{equation}
\boxed{
\begin{aligned}
\widehat{\boldsymbol{\beta}}_k^{\text{mon}} &= \argmin_{\boldsymbol{\beta}_k^{\text{mon}}\geq \mathbf{0}} \frac{1}{2}\left\|\mathbf{A}_{k,\mathbf{S}^{\text{non}}}\boldsymbol{\beta}_k^{\text{mon}}\right\|^2 - \sum_{i=1}^N \log\mathbf{b}_{k}\boldsymbol{\beta}_k^{\text{mon}} + \frac{1}{2}\boldsymbol{\beta}_k^{\text{mon},\intercal}\left(\mathbf{D}_{k,\mathbf{S}^{\text{non}}}^\intercal\mathbf{S}^{\text{non}}\mathbf{D}_{k,\mathbf{S}^{\text{non}}} + \mathbf{S}^{\text{mon}}\right)\boldsymbol{\beta}_k^{\text{mon}},\\ 
\widehat{\boldsymbol{\beta}}_k^{\text{non}} &= -\mathbf{M}_{k,\mathbf{S}^{\text{non}}}\mathbf{P}_{k}^{\text{mon}}\boldsymbol{\beta}_k^{\text{mon}}.
\end{aligned}}
\end{equation}

Note that the (element-wise) constraint of $\boldsymbol{\beta}_k^{\text{mon}}\geq \mathbf{0}$ is vital to maintain monotonicity; this can be enforced explicitly during optimization using particular optimization algorithms, e.g., L-BFGS-B (i.e. Low-memory BFGS with box constraints), implicitly by constructing an objective with a log-barrier term, or employing a convex transformation of the optimization objective (e.g. optimize over $\mathbf{p}_k^{\text{mon}} := \log\boldsymbol{\beta}_k^{\text{mon}}$). Since we often will have coefficients with zero values, the first methodology might be preferable (though no empirical results appear here). This optimization objective, notably, requires no evaluations of the map during optimization, that is to say, the optimization is as fast as the combination of the implementation of linear algebra algorithms called and the optimization routine used. %

\newcommand{\bfU}{\mathbf{U}_k}
\newcommand{\bfV}{\mathbf{V}_k}
\newcommand{\bfSig}{\boldsymbol{\Sigma}_k}

In the case where we have more parameters than samples, it is worth noting that using this formulation is remarkably sensitive to $\lambda$ because ${\mathbf{P}_{k}^{\text{non}}}^\top\mathbf{P}_{k}^{\text{non}}$ is no longer full rank. Thus the calculation of $\mathbf{M}_{k,\lambda}$ solves a system with possibly poor numerical properties (the industry of methods for \textit{ill-conditioned} systems is dedicated to such problems). For vanishingly small $\lambda$, this corresponds to the problem of overfitting and the fact that we have infinite choices of $\widehat{\boldsymbol{\beta}}_k^{\text{non}}$ for any given choice of $\boldsymbol{\beta}_k^{\text{mon}}$.

Finally, let us remark that for P-Splines, the monotonicity requirement does not only demand positive coefficients, but monotonously increasing coefficients. As such, let us define:

\begin{equation}
    \boldsymbol{\beta}_k^{\text{mon}} = \boldsymbol{T} \boldsymbol{\beta}_k^{\text{mon, nonincr.}}
\end{equation}

where $\boldsymbol{T}$ is a lower-triangular matrix of ones, creating a cumulative sum. Inserting this into the expression above yields:

\begin{equation}
\boxed{
\begin{aligned}
\widehat{\boldsymbol{\beta}}_k^{\text{mon}} &= \argmin_{\boldsymbol{T}\boldsymbol{\beta}_k^{\text{mon}}\geq \mathbf{0}} \frac{1}{2}\left\|\mathbf{A}_{k,\mathbf{S}^{\text{non}}}\boldsymbol{T}\boldsymbol{\beta}_k^{\text{mon}}\right\|^2 - \sum_{i=1}^N \log\mathbf{b}_{k}\boldsymbol{T}\boldsymbol{\beta}_k^{\text{mon}} + \frac{1}{2}\boldsymbol{\beta}_k^{\text{mon},\intercal}\boldsymbol{T}^\intercal\left(\mathbf{D}_{k,\mathbf{S}^{\text{non}}}^\intercal\mathbf{S}^{\text{non}}\mathbf{D}_{k,\mathbf{S}^{\text{non}}} + \mathbf{S}^{\text{mon}}\right)\boldsymbol{T}\boldsymbol{\beta}_k^{\text{mon}},\\ \widehat{\boldsymbol{\beta}}_k^{\text{non}} &= -\mathbf{M}_{k,\mathbf{S}^{\text{non}}}\mathbf{P}_{k}^{\text{mon}}\boldsymbol{T}\widehat{\boldsymbol{\beta}}_k^{\text{mon}}.
\end{aligned}}
\end{equation}

And where $\mathbf{A}_{k,\mathbf{S}^{\text{non}}}$ is defined as:

\begin{equation}
\begin{aligned}
    \mathbf{A}_{k,\mathbf{S}^{\text{non}}} &= \left(\mathbf{I} - \mathbf{P}_{k}^{\text{non}}\mathbf{M}_{k,\mathbf{S}^{\text{non}}}\right)\mathbf{P}_{k}^{\text{mon}} \\
    &= \left(\mathbf{I} - \mathbf{P}_{k}^{\text{non}}\left({\mathbf{P}_{k}^{\text{non}}}^{\top}\mathbf{P}_{k}^{\text{non}}+ \mathbf{S}^{\text{non}}\right)^{-1}{\mathbf{P}_{k}^{\text{non}}}^\top\right)\mathbf{P}_{k}^{\text{mon}} \\
\end{aligned}
\end{equation}

Likewise, $\mathbf{D}_{k,\mathbf{S}^{\text{non}}}$ is defined as:

\begin{equation}
\begin{aligned}
    \mathbf{D}_{k,\mathbf{S}^{\text{non}}} &= \mathbf{M}_{k,\mathbf{S}^{\text{non}}}\mathbf{P}_{k}^{\text{mon}} \\
    &= \left({\mathbf{P}_{k}^{\text{non}}}^{\top}\mathbf{P}_{k}^{\text{non}}+ \mathbf{S}^{\text{non}}\right)^{-1}{\mathbf{P}_{k}^{\text{non}}}^\top\mathbf{P}_{k}^{\text{mon}} \\
\end{aligned}
\end{equation}

\section{P-splines}
\label{appendix:p-splines}

In this section, we discuss the concept of \textbf{penalized basis splines} (P-splines), and why they are an attractive choice for the parametrization of adaptive triangular transport maps.

\subsection{Basis splines}

\begin{figure}[ht!]
  \centering
  \includegraphics[width=\textwidth]{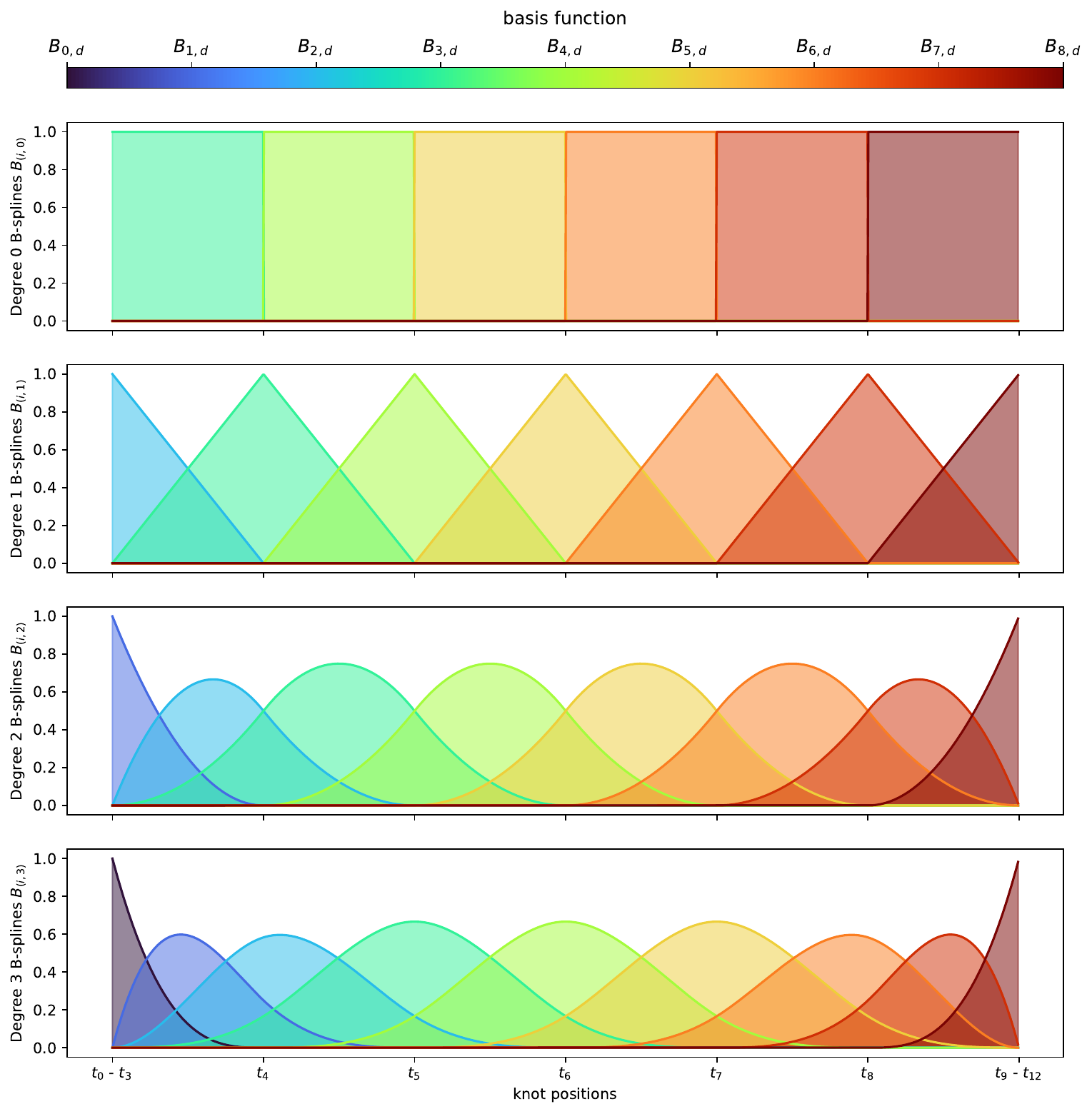}
  \caption{The four subplots illustrate the basis function evaluations for (from top to bottom) degree zero, degree one, degree two, and degree three for a B-spline with seven "real" knots. Observe that each spline becomes smoother and broader as the recursion iterates through the degrees. Each higher degree adds a "new" basis function to the left as the interpolation intervals widen. These new basis functions also exist at lower degrees, but are degenerate / zero, since they only correspond to degenerate intervals between padding knots. Only once the interpolation intervals in Equation~\ref{eq:B_spline_higher} includes real knots do these functions become non-zero.}
  \label{fig:P_splines_degrees}
\end{figure}

To understand P-Splines, we should first discuss basis splines, or B-splines. B-splines are basis functions with limited support, and are often created with the recursive \textbf{de Boor algorithm} \citep{de_boor_practical_2001}, which constructs higher-order splines by interpolating lower-order splines. These splines are each positioned between a series of $m+1$ knots $t_{0},\dots,t_{m}$. 

To begin, degree zero splines ($d=0$) are defined as indicator functions for the interval between two adjacent knots

\begin{equation}
    B_{i,d=0}(x) = \begin{cases}
        1 & \text{if } t_{i} \leq x \leq t_{i+1} \\
        0 & \text{otherwise}\\
    \end{cases}
    \label{eq:B_spline_zero}
\end{equation}

for all $i=0,\dots,m-1$. Higher-degree splines are subsequently derived in a recursion that interpolates B-splines of one degree lower:

\begin{equation}
    B_{i,d>0}(x) = \frac{x - t_i}{t_{i+d}-t_{i}} B_{i,d-1}(x) + \frac{t_{i+d+1} - x}{t_{i+d+1}-t_{i+1}} B_{i+1,d-1}(x).
    \label{eq:B_spline_higher}
\end{equation}

Note that at degree $d=0$, the B-spline $B_{i,d=0}$ is only non-zero between knots $t_{i}$ and $t_{i+1}$. At degree $d=1$, the spline now interpolates between knots $t_{i}$ and $t_{i+d+1}=t_{i+2}$, so we now interpolate between two knot. Observe that now a B-spline at a hypothetical index $B_{i=-1,d=1}$ to the left of the original knots could be non-zero, as it would have a non-zero second term in Equation~\ref{eq:B_spline_higher}. Likewise, the basis spline at the last "real" knot $t_{m}$ requires additional knots $t_{m+1},\dots,t_{m+d+1}$ to the right for higher-degree evaluations of Equation~\ref{eq:B_spline_higher}. An illustration of B-splines at different degrees is provided in Figure~\ref{fig:P_splines_degrees}.

In consequence, it is common practice to "pad" the knots with number of degrees $d$ extra knots on both sides. For instance, for cubic B-splines ($d=3$) with real knots at locations $(0,1,2,3)$, the extended knots would be defined as

\begin{equation}
    \begin{aligned}
    \text{knot} \quad &&t_{0} && t_{1} && t_{2} && t_{3} && t_{4} && t_{5} && t_{6} && t_{7} && t_{8} && t_{9} \\
    \text{position} \quad && 0 && 0 && 0 && 0 && 1 && 2 && 3 && 3 && 3 && 3 \\
    \text{type} \quad &&\text{pad} && \text{pad} && \text{pad} && \text{real} && \text{real} && \text{real} && \text{real} && \text{pad} && \text{pad} && \text{pad}
    \end{aligned},
    \label{eq:B_spline_knots}
\end{equation}

where we observe that $x$ values can only fall into intervals between "real" knots, but the padded knots can be used to compute B-Splines for indices $i=0,\dots,m+d$; in the example in Equation~\ref{eq:B_spline_knots}, we can define B-splines up to index $i_{\max}=m+d$, that is to say, up to B-spline $B_{i_{\max},d=3}$:

\begin{equation}
    f(x; \boldsymbol{\beta}) = \sum_{i=0}^{m+d} \beta_{i} B_{i,d}(x).
    \label{eq:B_spline_approximation}
\end{equation}

We note that other strategies to construct B-splines exist, and refer the interested reader to the Appendix of \citet{eilers_practical_2021} for a discussion of alternative approaches. The penalty part of P-splines introduces a penalty on higher-order.

\subsection{Linearized B-splines}\label{subsubsec:lin_b_splines}

While P-splines are a robust and flexible class of functions with the support of their knots, a computationally robust definition of triangular maps also requires accounting for evaluations beyond the assigned knots. In general, linear extrapolation is a robust choice when extrapolating map component functions \citep[e.g., ][Section 3.1]{ramgraber_friendly_2025}. Other concerns in the design of P-splines include how many "real" knots should be used, and where they should be placed.

In this manuscript, we propose the following strategy to define robust linearly-extrapolated P-splines:
\begin{enumerate}
    \item \textbf{Number of knots}. 
    Following the discussion in Section~\ref{subsec: computational considerations}, the number of knots should be large enough to avoid restricting model flexibility, while remaining computationally manageable. We therefore set the number of interior knots for each marginal P-spline to
    \begin{align}
        K = \operatorname{ceil}(n^{1/3}) + 2,
    \end{align}
    where $n$ denotes the number of (unique) training samples. The additional two knots ensure stable linear extrapolation in the tails.
    \item \textbf{Positions of knots}. We place each real knot at equal spacings between the empirical $10\%$ and $90\%$ quantiles along the corresponding dimension for each marginal P-spline, such that $t_{\text{first}}=t_{d}=q_{10}(x)$ and $t_{\text{last}}=t_{d+m}=q_{90}(x)$ are the first and last real knots, respectively.
    \item \textbf{Linear extrapolation}. When a spline is evaluated outside the range of the real knots, we extrapolate the splines linearly:
\end{enumerate}
\begin{equation}
    f(x;\boldsymbol{\beta}) = \begin{cases}
        \sum_{i=0}^{m+d} \beta_{i} B_{i,d}(t_{\text{first}}) + (x - t_{\text{first}}) \sum_{i=0}^{m+d} \beta_{i} \frac{\partial B_{i,d}(t_{\text{first}})}{\partial x}  & \text{if } x < t_{\text{first}}  \\
        \sum_{i=0}^{m+d} \beta_{i} B_{i,d}(x) & \text{if } t_{\text{first}} \leq x \leq t_{\text{last}} \\
        \sum_{i=0}^{m+d} \beta_{i} B_{i,d}(t_{\text{last}}) + (x - t_{\text{last}}) \sum_{i=0}^{m+d} \beta_{i} \frac{\partial B_{i,d}(t_{\text{last}})}{\partial x}  & \text{if } t_{\text{last}} < x \\
    \end{cases}.
\end{equation}

\subsection{Smoothness penalty}

\begin{figure}[ht!]
  \centering
  \includegraphics[width=\textwidth]{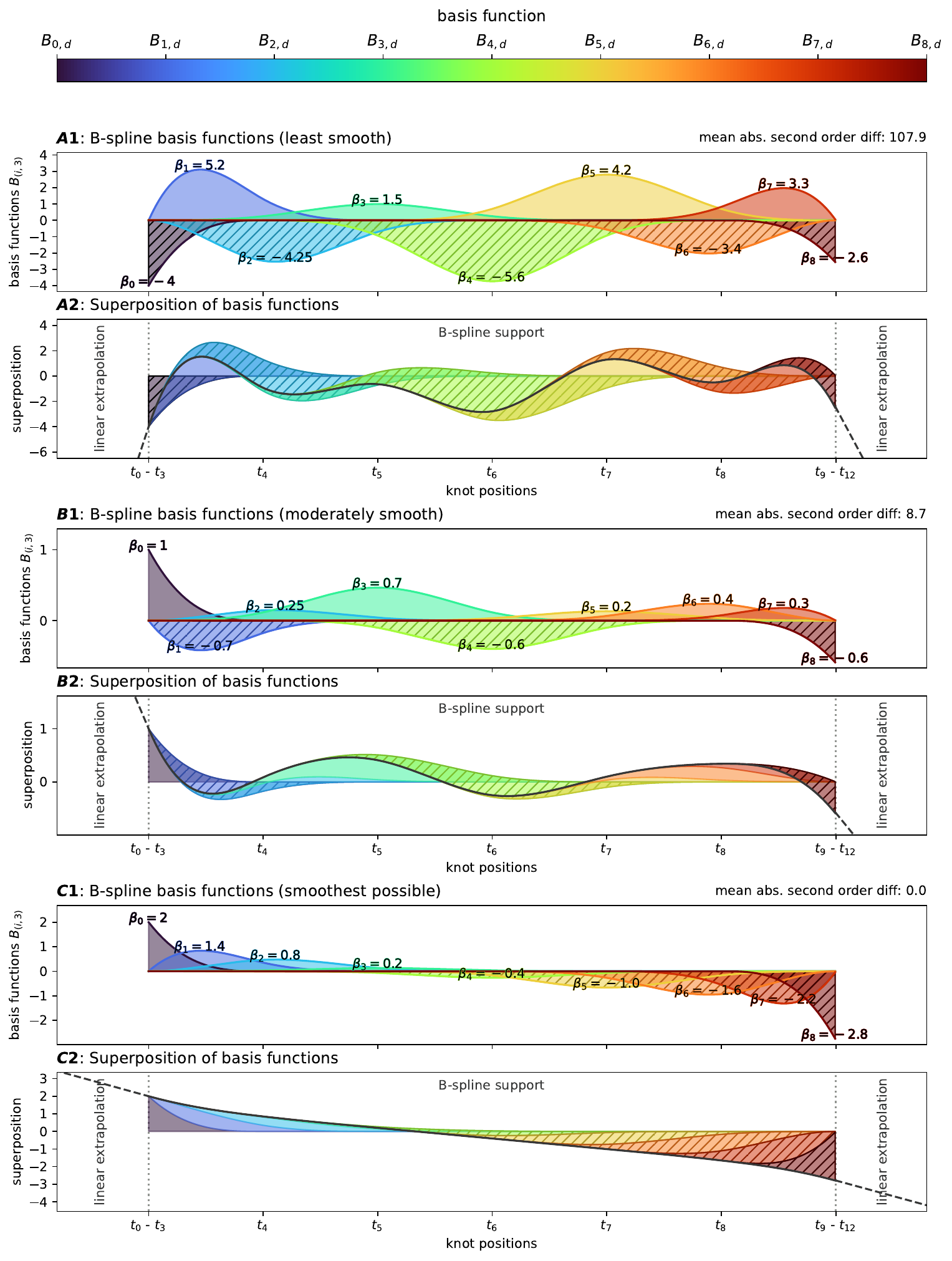}
  \caption{Reducing the second order differences between coefficients $\beta_{i}$ at adjacent B-spline basis functions results in smoother superpositions. Subplots from A to C show P-splines with decreasing second-order coefficient differences. Hatched regions denote B-spline basis functions with negative coefficients.}
  \label{fig:B_splines_smooth}
\end{figure}

The eponymous penalization term enters the P-spline formulation as a penalty on higher-order differences for the coefficients $\boldsymbol{\beta}=[\beta_{0},\dots,\beta_{m+d}]$ of adjacent B-spline basis functions. For degree $d=3$ P-splines, it is common to penalize second-order differences between the coefficients.

\FloatBarrier %

\bibliographystyle{unsrtnat}
\bibliography{references}  %

\end{document}